\documentclass[12pt]{article}
\usepackage{jheppub}
\pdfoutput=1
\usepackage{epstopdf}
\usepackage{slashed}
\usepackage{mathrsfs}
\usepackage{psfrag}
\usepackage{epsf}
\usepackage{graphicx,epsfig}
\usepackage{amssymb}
\usepackage{epstopdf}
\usepackage{float}
\usepackage{braket}
\usepackage{hyperref}
\usepackage{xcolor}
\usepackage{mathtools}
\usepackage{physics}
\usepackage{tikz-feynman}
\tikzfeynmanset{compat=1.0.0}
\usepackage{amsmath,array,amsfonts,wrapfig,lscape,multirow,longtable,rotating,subfigure,makecell}
\usepackage{url}

\newcolumntype{C}[1]{>{\centering\arraybackslash}m{#1}}
\newcommand{\ClaudeComment}[1]{{\bf [CD: #1]}}
\newcommand{\rayan}[1]{{\bf [RH: #1]}}
\newcommand{\achilleas}[1]{{\bf [AL: #1]}}

\newcommand{\K}[1]{k_{\perp #1}}
\newcommand{\Kb}[1]{\kappa_{\perp #1}}
\newcommand{\KT}[1]{\tilde{k}_{\perp #1}}
\newcommand{\KTb}[1]{\tilde{\kappa}_{\perp #1}}
\newcommand{\pM}[1]{\widetilde{P}^{#1}}
\newcommand{\pN}[1]{\widetilde{P}'^{#1}}

\def\beq{\begin{equation}}
\def\eeq{\end{equation}}

\newcolumntype{C}[1]{>{\centering\arraybackslash}m{#1}}

\newcommand{\comments}[1]{}

\def\cA{{\cal A}}  \def\cC{{\cal C}}

\def\cM{{\cal M}}  
  
\def\cS{{\cal S}} \def\cT{{\cal T}} \def\cU{{\cal U}}

\def\eps{\epsilon}

\graphicspath{{./images/}}

\def\beq{\begin{equation}}
\def\eeq{\end{equation}}
\def\bsp#1\esp{\begin{split}#1\end{split}}
\def\cM{{\cal M}}

\preprint{CERN-TH-2019-221, CP3-19-57}

\title{Tree-level splitting amplitudes for a quark into four collinear partons}

\author[a]{{\small Vittorio Del Duca}\footnote{On leave from INFN, Laboratori Nazionali di Frascati, Italy.}}
\author[b]{{\small Claude Duhr}}
\author[a]{{\small Rayan Haindl}}
\author[a]{{\small Achilleas Lazopoulos}}
\author[c]{{\small Martin Michel}}

\affiliation[a]{Institute for Theoretical Physics, ETH Z\"{u}rich, 8093 Z\"{u}rich, Switzerland.}
\affiliation[b]{Theoretical Physics Department, CERN, CH-1211 Geneva 23, Switzerland.}
\affiliation[c]{Center for Cosmology, Particle Physics and Phenomenology (CP3),\\
UCLouvain, B-1348, Louvain-La-Neuve, Belgium.}

\emailAdd{delducav@itp.phys.ethz.ch, claude.duhr@cern.ch, haindlr@phys.ethz.ch, lazopoli@phys.ethz.ch, martin.michel@uclouvain.be}

\abstract{
We compute in conventional dimensional regularisation
the tree-level splitting amplitudes for a quark parent
in the limit where four partons become collinear to each other. This is part of the universal infrared behaviour of the QCD scattering amplitudes at 
next-to-next-to-next-to-leading order (${\rm N^3LO}$) in the strong coupling constant.
Further, we consider
the iterated limit when $m'$ massless partons become collinear to each other within a bigger set of $m$ collinear partons,
as well as the limits when one gluon or a $q\bar{q}$ pair or two gluons become soft within a set of $m$ collinear partons.
}

\begin{document}

\maketitle

\section{Introduction}
\label{sec:intro}

The measurements of the Higgs boson properties~\cite{Dittmaier:2011ti,Dittmaier:2012vm,Heinemeyer:2013tqa,deFlorian:2016spz}
 are a paradigm of the physics program at the Large Hadron Collider (LHC). A precise determination of the Higgs couplings
is crucial to test possible deviations from the Standard Model. 

In hadron collisions, 
the largest production mode of the Higgs boson
is from gluon fusion. The Higgs boson couples to the gluons via a heavy-quark loop. Accordingly, the inclusive Higgs production cross section
is known at next-to-leading order (NLO)~\cite{Graudenz:1992pv,Spira:1995rr} in the strong coupling constant $\alpha_s$ of perturbative QCD, with full heavy-quark mass dependence. The NLO corrections more than double the leading-order cross section. In this fashion, the next-to-next-to-leading order (NNLO) corrections are not known yet.
However, in the limit in which the heavy-quark mass is much larger than the other scales in the process, 
i.e. by replacing the loop-mediated Higgs-gluon coupling by a tree-level effective coupling, 
which is often termed Higgs Effective Field Theory (HEFT), inclusive Higgs production from gluon fusion is known at 
next-to-next-to-next-to-leading order (${\rm N^3LO}$)~\cite{Anastasiou:2015ema,Mistlberger:2018etf}, whose accuracy has reached the 5\% level~\cite{Anastasiou:2016cez}. 
The next-to-largest production mode of the Higgs boson
in hadron collisions is from vector-boson fusion, for
which inclusive Higgs~\cite{Dreyer:2016oyx} and double-Higgs~\cite{Dreyer:2018qbw} production are known at ${\rm N^3LO}$ in the deep inelastic scattering (DIS) approximation. Also the cross section for the production of a Higgs boson in bottom quark fusion in the five flavour scheme is known at N$^3$LO in $\alpha_s$~\cite{Duhr:2019kwi}.

To go beyond inclusive cross sections with hadronic initial states requires to measure and compute differential distributions. 
Only very few are known at ${\rm N^3LO}$ accuracy: the Higgs rapidity distribution, computed in the HEFT~\cite{Cieri:2018oms,Dulat:2018bfe}, the Higgs rapidity and transverse momentum distributions in vector boson fusion in the DIS approximation~\cite{Dreyer:2016oyx}, and
jet production in DIS~\cite{Currie:2018fgr}.
More differential distributions, like the Higgs transverse momentum distribution~\cite{Chen:2016zka} 
or Higgs production in association with a jet~\cite{Boughezal:2015dra,Boughezal:2015aha},
are known in the HEFT at NNLO in $\alpha_s$.
They usually involve (two-loop) $2\to 2$ scattering amplitudes, and sector-decomposition~\cite{Binoth:2000ps,Anastasiou:2005cb},
slicing~\cite{Catani:2007vq,Boughezal:2015eha,Gaunt:2015pea} or
subtraction~\cite{
GehrmannDeRidder:2005cm,Daleo:2006xa,GehrmannDeRidder:2005aw,
GehrmannDeRidder:2005hi,Daleo:2009yj,Gehrmann:2011wi,Boughezal:2010mc,GehrmannDeRidder:2012ja,Currie:2013vh,
Somogyi:2005xz,Somogyi:2006da,
Somogyi:2006db,Somogyi:2008fc,Aglietti:2008fe,Somogyi:2009ri,
Bolzoni:2009ye,Bolzoni:2010bt,DelDuca:2013kw,Somogyi:2013yk, 
Czakon:2010td,Czakon:2011ve,Czakon:2014oma,Czakon:2019tmo, 
Caola:2017dug,Caola:2018pxp,Delto:2019asp,Caola:2019nzf,Caola:2019pfz,Cacciari:2015jma}
methods at NNLO,\footnote{Projection to Born~\cite{Cacciari:2015jma}
and $q_T$ subtraction~\cite{Catani:2007vq} have been successfully used also in evaluations at ${\rm N^3LO}$ accuracy, 
Refs.~\cite{Dreyer:2016oyx,Currie:2018fgr,Dreyer:2018qbw} and \cite{Cieri:2018oms}, respectively.} to deal with the computation of
the cross section where the phase-space integration of the (double and single) real radiation is performed with generic acceptance cuts.
With some abuse of terminology, we shall generically refer to those methods as \emph{subtraction methods}.

Despite the recent progress in computing cross sections and differential distributions at N$^3$LO in the strong coupling constant, extending these results to more general observables and processes requires further developments on the theory side, both for the computation of the purely virtual three-loop corrections and for the development of subtraction methods at N$^3$LO. Indeed, on the one hand, so far there is no complete 2-to-2 three-loop scattering amplitude known in QCD -- only the three-loop
gluon-gluon scattering amplitude in $\mathcal{N}=4$ super-Yang-Mills theory is known~\cite{Henn:2016jdu} (though the structure of the infrared singularities is known for generic massless three-loop amplitudes~\cite{Almelid:2015jia,Almelid:2017qju}).
On the other hand,  the tenet of the subtraction methods is the universal behaviour of the QCD scattering amplitudes in the infrared -- soft and/or collinear -- limits.
Understanding the behaviour of the amplitudes in these limits at a given order is therefore a necessary ingredient to develop a subtraction scheme at that order.

At NNLO, the infrared behaviour of amplitudes is embodied in the tree-level currents, with three collinear partons or two soft
partons~\cite{Campbell:1997hg,Catani:1999ss,DelDuca:1999iql,Kosower:2002su}; and in the one-loop currents, with two collinear partons 
or one soft gluon~\cite{Bern:1994zx,Bern:1998sc,Kosower:1999rx,Bern:1999ry,Catani:2000pi}.
The NNLO counterterms for unresolved double-parton and single-parton configurations are then built out of those universal infrared currents.

The infrared behaviour of amplitudes at ${\rm N^3LO}$ is embodied in the two-loop currents, with two collinear
partons~\cite{Bern:2004cz,Badger:2004uk,Duhr:2014nda} or one soft gluon~\cite{Duhr:2013msa,Li:2013lsa}; 
in the one-loop currents, with three collinear partons~\cite{Catani:2003vu,Badger:2015cxa}\footnote{One-loop currents with three collinear partons are also known for
mixed QCD+QED cases~\cite{Sborlini:2014mpa}.} or two soft partons; and in the tree-level currents, 
with four\footnote{Tree-level currents with more than four collinear gluons are known for MHV configurations~\cite{Birthwright:2005ak}.} collinear partons~\cite{DelDuca:1999iql,Birthwright:2005ak,Duhr:2006} or three soft partons~\cite{Catani:2019nqv}.\footnote{In Ref.~\cite{Catani:2019nqv}, the triple
soft-gluon radiation is displayed, including some results on the quadruple soft-gluon radiation from two hard partons.}

The results for the splitting amplitudes of Refs.~\cite{DelDuca:1999iql,Birthwright:2005ak,Duhr:2006}, however, are not directly applicable 
to the construction of a subtraction method, because they have been extracted from four-dimensional amplitudes with fixed external helicity states. In order to regulate the phase space divergences it is important to work with a suitable regularisation scheme. A particularly convenient such scheme is the conventional dimensional regularisation (CDR) scheme, for which the number of spacetime dimensions is fixed to $D=4-2\eps$, and the quarks and gluons
have 2 and $D-2$ helicity states, respectively.
In this paper we compute the squared tree-level quadruple-collinear splitting amplitudes for a quark parent in the CDR scheme. Further, we compute all ensuing iterated limits where some of the collinear partons become themselves unresolved, i.e., soft or collinear. Indeed, in designing counterterms for subtraction methods beyond NLO, it has become evident that it is important to have a detailed knowledge of iterated limits which describe overlapping and strongly-ordered divergences.

This paper is organised as follows:
In Sec.~\ref{sec:setup}, we discuss in general the limit when $m$ massless partons become collinear to each other
in a tree-level amplitude.
In Sec.~\ref{sec:collinear}, we present the main result of our paper, namely the computation of the tree-level splitting amplitudes for a quark parent to split into four collinear partons. The explicit results, which are too long to be recorded in this paper, are made available in computer-readable form~\cite{QuadColKernelsWebsite}.  In Sec.~\ref{sec:nested_collinear} we discuss in general the limit when $m'$ massless partons become collinear to each other within a bigger cluster of $m$ collinear partons.
We then specify the cases when two or three partons become collinear to each other within a cluster of four collinear partons. 
In Sec.~\ref{sec:iterated_soft-collinear}, we review
the tree-level soft currents that we use in this paper,
namely the single soft gluon and the double soft currents at tree-level.
In Sec.~\ref{sec:singlesoftcollamp}, we describe
the limit when one gluon becomes soft within a
cluster of $m$ collinear partons, and then we consider the
particular cases of $m=2, 3, 4$.
In Sec.~\ref{sec:doublesoftlim}, we describe
the limit when a $q\bar{q}$ pair or two gluons become soft within a cluster of $m$ collinear partons, specifying it then to $m=3, 4$.
In Sec.~\ref{sec:conclusion}, we summarise our findings. We include several appendices with technical material omitted throughout the main text.

\section{Multiple collinear limits at tree level}
\label{sec:setup}

The aim of this paper is to study the behaviour of tree-level QCD amplitudes in the limit where a certain number of massless partons become collinear. To be more concrete, consider a scattering of $n$ particles with momenta $p_i$ and with flavour, spin and colour quantum numbers $f_i$, $s_i$ and $c_i$, respectively, and assume that $p_i^2=0$ for $1\le i\le m\le n-3$. Our goal is to study the behaviour of the amplitude as $p_1,\ldots,p_m$ become simultaneously collinear to some light-like direction $\pM{}$. It is well known that in this limit the scattering amplitude diverges order by order in perturbation theory, and the leading behaviour in the limit is described by the amplitude for the production of a massless particle of momentum $\pM{}$ from a scattering of the particles that do not take part in the collinear limit, multiplied by a universal factor which captures the collinear divergence and only depends on the particles in the collinear set. In the remainder of this section we define the collinear limit we are interested in more precisely and we set our notations and conventions.

In order to parametrise the approach to the collinear limit, it is convenient to introduce a light-cone decomposition for all the momenta in the $m$-parton collinear set,
\begin{equation} 
\label{eq:collinear_momenta}
p_i^\mu = x_i\pM{\mu} + \K{i}^\mu - \frac{\K{i}^2}{2 x_i}\frac{n^\mu}{\pM{}\cdot n}\,,
\qquad
i = 1,\ldots, m\,,
\end{equation}
where the light-like momentum $\pM{}$ specifies the collinear direction, 
$\pM{} \cdot k_{\perp i} = 0$, $x_i$ are the longitudinal momentum fractions with respect to the parent momentum $P^\mu = \sum_{i=1}^m p_i^\mu$ and
$n^\mu$ is an auxiliary light-like vector such that $n\cdot\K{i} = 0$ and $n\cdot p_{i} \neq 0\neq n\cdot\pM{}$, and
which specifies how the collinear direction is approached. The collinear limit is then defined as the limit in which the transverse momenta $\K{i}$ approach zero at the same rate. We stress that this definition of the collinear limit is frame-independent, and it only depends on the collinear direction $\pM{}$ and the transverse momenta $\K{i}$. In particular it is independent of the choice of the auxiliary vector $n$; changing the direction of $n$ will merely affect the direction from which the collinear limit is approached, but not the behaviour of the amplitude in the limit. 

The variables that appear in eq.~\eqref{eq:collinear_momenta} are, at least a priori, completely unconstrained apart from on-shellness and transversality, $n\cdot \K{i} = \pM{}\cdot \K{i} = 0$, and so the sums of the momentum fractions $x_i$ and the transverse momenta $\K{i}$ are unconstrained.
Therefore, the parametrisation in eq.~\eqref{eq:collinear_momenta} seems to depend on $(D-1)m+2$ degrees of freedom in $D$ space-time dimensions: the $m$ variables $x_i$ and $\K{i}$ in addition to the two light-like directions $\pM{}$ and $n$. This naive counting seems to be at odds with the fact that a set of $m$ light-like momenta (that do not sum up to zero) depend on $(D-1)m$ degrees of freedom. This apparent conundrum is resolved upon noting that the collinear limit is invariant under longitudinal boosts in the direction of the parent momentum 
$P = \sum_{i =1}^m p_i$. This suggests to trade $x_i$ and $\K{i}$ for new quantities $z_i$ and $\KT{i}$ that are boost-invariant in the direction of the parent momentum. In App.~\ref{app:boost} we show that a convenient set of such variables is given by
\begin{equation}
z_i = \frac{x_i}{\sum_{j=1}^m x_j} = \frac{p_i\cdot n}{P\cdot n}\,, 
\qquad \KT{i}^\mu = \K{i}^{\mu}- z_i\sum_{j=1}^m \K{j}^{\mu}\,
\qquad
i = 1,\ldots, m\,.
\label{eq:boost}
\end{equation}
It is easy to see that these new variables satisfy the constraints,
\beq
\sum_{i=1}^m z_i =1 {\rm~~and~~} \sum_{i=1}^m\KT{i}^\mu = 0\,,
\label{eq:constraint}
\eeq
thereby reducing the number of degrees of freedom to $(D-1)m$.

It is well known that in the limit where a subset of massless particles is collinear a scattering amplitude factorises as~\cite{Amati:1978wx,Amati:1978by,Ellis:1978sf}
\beq\bsp\label{eq:split_def}
\mathscr{C}_{1\ldots m}\cM&_{f_1\ldots f_n}^{c_1\ldots c_n;s_1\ldots s_n}(p_1,\ldots,p_n)\\
&\, = {\bf Sp}_{ff_1\ldots f_m}^{c,c_1\ldots c_m;s,s_1\ldots s_m}\,\cM_{ff_{m+1}\ldots f_n}^{c,c_{m+1}\ldots c_n;s,s_{m+1}\ldots s_n}(\pM{},p_{m+1},\ldots,p_n)\,.
\esp\eeq
Here $\mathscr{C}_{1\ldots m}$ indicates that the equality only holds up to terms that are power-suppressed in the collinear limit, while $f$, $s$ and $c$ respectively denote the flavour, spin and colour indices of the parent particle. Note that in a theory with only fermions and gluons the flavour of the parent is uniquely determined by the flavours of the particles in the collinear set. We will therefore often suppress the dependence of the splitting amplitude on the flavour of the parent parton. The quantity ${\bf Sp}$ appearing on the right-hand side is called a splitting amplitude and only depends on the kinematics and the quantum numbers in the collinear set. 

The factorisation in eq.~\eqref{eq:split_def} is valid to all orders in perturbation theory.
It implies that also the squared matrix element must factorise. We use the following notation for the matrix element summed over all spin and colour indices,
\beq
\left|\cM_{f_1\ldots f_n}(p_1,\ldots,p_n)\right|^2 \equiv \sum_{\substack{(s_1,\ldots,s_n)\\(c_1,\ldots,c_n)}}\left|\cM_{f_1\ldots f_n}^{c_1\ldots c_n;s_1\ldots s_n}(p_1,\ldots,p_n)\right|^2\,.
\eeq
Using this notation, the factorisation of the squared matrix element can be concisely written as
\beq
\mathscr{C}_{1\ldots m}\left|\cM_{f_1\ldots f_n}(p_1,\ldots,p_n)\right|^2 = \Bigg(\frac{2\mu^{2\epsilon}\,g_s^2}{s_{1\ldots m}}\Bigg)^{m-1} \,\hat{P}^{ss'}_{f_1\ldots f_m}\, \mathcal{T}_{ff_{m+1}\ldots f_n}^{ss'}(\pM{},p_{m+1},\ldots,p_n)\,,
\label{eq:factorisation}
\eeq
where $g_s$ denotes the strong coupling constant and $\mu$ is the scale introduced by dimensional regularisation. A sum over the spin indices $s$ and $s'$ is understood, and we defined the Mandelstam invariant
\beq
s_{1\ldots m} \equiv (p_1+\ldots+p_m)^2\,.
\eeq
Throughout this paper we work in Conventional Dimensional Regularisation (CDR), where the gluons have $D-2$ polarisations. $\mathcal{T}_{ff_{m+1}\ldots f_n}^{ss'}$ denotes the helicity tensor obtained by not summing over the spin indices of the parent parton,
\beq\label{eq:T_def}
 \mathcal{T}_{ff_{m+1}\ldots f_n}^{ss'} \equiv \sum_{\substack{(s_{m+1},\ldots,s_n)\\(c,c_{m+1},\ldots,c_n)}}\cM_{ff_{m+1}\ldots f_n}^{c,c_{m+1}\ldots c_n;s,s_{m+1}\ldots s_n}\,\left[\cM_{ff_{m+1}\ldots f_n}^{c,c_{m+1}\ldots c_n;s',s_{m+1}\ldots s_n}\right]^{\ast}\,,
\eeq
where for the sake of clarity we have suppressed the momenta on which the amplitude depends. The tensorial structure of the factorisation in eq.~\eqref{eq:factorisation} is necessary to correctly capture all spin correlations. Due to colour conservation in the hard amplitude there are no non-trivial colour correlations, and we therefore sum over the colour $c$ of the parent parton in eq.~\eqref{eq:T_def}. The quantity $\hat{P}^{ss'}_{f_1\ldots f_m}$ in eq.~\eqref{eq:factorisation} is the (polarised) splitting amplitude for the squared matrix element. It depends on the transverse momenta $\KT{i}$ and momentum fractions $z_i$ of the particles in the collinear set as well as the spin indices of the parent parton. As in general no confusion will arise, we will always suppress the dependence of the splitting amplitude on its arguments. It is related to ${\bf Sp}$ by 
\beq\label{eq:P_def}
\Bigg(\frac{2\mu^{2\epsilon}\,g_s^2}{s_{1\ldots m}}\Bigg)^{m-1} \,\hat{P}^{ss'}_{f_1\ldots f_m} =
\frac1{\mathcal{C}_{f}}
\sum_{\substack{(s_{1},\ldots,s_m)\\(c,c_{1},\ldots,c_m)}}  {\bf Sp}_{ff_1\ldots f_m}^{c,c_1\ldots c_m;s,s_1\ldots s_m}\,\left[{\bf Sp}_{ff_1\ldots f_m}^{c,c_1\ldots c_m;s',s_1\ldots s_m}\right]^{\ast}\,,
\eeq
where $\mathcal{C}_{f}$ is the number of colour degrees of freedom of the parent parton with flavour $f$, i.e., $\mathcal{C}_{g}=N_c^2-1$ for a gluon and $\mathcal{C}_q=N_c$ for a quark.

It is sometimes useful to define the unpolarised splitting amplitude by averaging over the spins of the parent parton,
\beq
\langle\hat{P}_{f_1\ldots f_m}\rangle \equiv \frac{1}{N_{\textrm{pol}}}\,\delta_{ss'}\,\hat{P}^{ss'}_{f_1\ldots f_m}\,,
\eeq
where $N_{\textrm{pol}}$ denotes the number of physical polarisation states for the parent parton.
The splitting amplitudes for the squared matrix element have been computed at tree level for the emission of up to three collinear partons in ref.~\cite{Campbell:1997hg,Catani:1999ss}. The goal of this paper is to compute for the first time the tree-level splitting amplitudes\footnote{Since in general no confusion arise, we refer to both ${\bf Sp}$ and $\hat{P}$ simply as \emph{splitting amplitudes}.} for the squared matrix element for the emission of up to four partons, in the case where the parent parton is a quark.

When the parent parton is a quark, Lorentz invariance implies that the fermion number and the helicity must be conserved. Then it is easy to see that the tensorial structure of the splitting amplitude is trivial,
\beq
\hat{P}^{ss'}_{f_1\ldots,f_m} = \delta^{ss'}\,\langle \hat{P}_{f_1\ldots f_m}\rangle\,.
\label{eq:unpol}
\eeq
For $m=2$ and $m=3$ the corresponding splitting amplitudes read~\cite{Altarelli:1977zs,Catani:1999ss},
\begin{align}
\label{eq:P23}
\langle \hat{P}_{qg}\rangle &\,=C_F\left( \frac{1+z^2}{1-z}-\eps (1-z)\right)\,,\\
\nonumber\langle \hat{P}_{\bar{q}_1'q_2'q_3}\rangle &\,=C_FT_R Q_{12,3}\,,\\
\nonumber\langle \hat{P}_{\bar{q}_1q_2q_3}\rangle &\,=C_F T_R \left( Q_{12,3} + Q_{13,2}\right) + C_F\left(C_F-\frac{C_A}{2}\right)\left(Q^{(id)}_{12,3}+Q^{(id)}_{13,2}\right)\,,\\
\nonumber\langle \hat{P}_{g_1g_2q_3}\rangle &\,=C_F^2 \left(Q^{(ab)}_{12,3} + Q^{(ab)}_{21,3}\right) + C_FC_A \left(Q^{(nab)}_{12,3} + Q^{(nab)}_{21,3} \right) \,,
\end{align}
with
\beq\label{eq:nnlo_splitting_kernels_1}
Q_{ij,k} = \frac{1}{2} \frac{s_{ijk}}{s_{ij}}\left[ -\frac{t_{ij,k}^2}{s_{ij}s_{ijk}}+\frac{4z_k+(z_i-z_j)^2}{z_i+z_j}+(1-2 \eps)\left( z_i+z_j-\frac{s_{ij}}{s_{ijk}} \right) \right]\,,
\eeq
\begin{align}\label{eq:nnlo_splitting_kernels_2}
Q^{(id)}_{ij,k} &= 
(1-\epsilon)\left(\frac{2s_{jk}}{s_{ij}} - \epsilon\right) \\
\nonumber&
+\frac{s_{ijk}}{s_{ij}}\left[  \frac{1+z_i^2}{1-z_j} - \frac{2z_j}{1-z_k} -\epsilon \left(\frac{(1-z_k)^2}{1-z_j} + 1 + z_i -2\frac{z_j}{1-z_k}\right) - \epsilon^2(1-z_k) \right]
\\
\nonumber&
-\frac{s_{ijk}^2}{s_{ij}s_{ik}}\frac{z_i}{2}\left[ \frac{1+z_i^2}{(1-z_j)(1-z_k)}-\epsilon \left(1+2\frac{1-z_j}{1-z_k}\right)
-\epsilon^2
\right] \,,
\end{align}
\begin{align}\label{eq:nnlo_splitting_kernels_3}
Q^{(ab)}_{ij,k} &= \frac{s_{ijk}^2}{2 s_{ik}
   s_{jk}} z_k \left[\frac{z_k^2+1}{z_i z_j}-\frac{\epsilon 
   \left(z_i^2+z_j^2\right)}{z_i z_j}-\epsilon  (\epsilon +1)\right]
   \\
 \nonumber  &+\frac{s_{ijk}}{s_{ik}} \left[\frac{\left(1-z_i\right)
   z_k+\left(1-z_j\right){}^3}{z_i z_j}-\frac{\epsilon  \left(1-z_j\right) \left(z_i
   z_j+z_i^2+z_j^2\right)}{z_i z_j}+\epsilon ^2
   \left(z_k+1\right)\right]
   \\
 \nonumber  &
   +(1-\epsilon ) \left[\epsilon -(1-\epsilon )\frac{
   s_{jk}}{s_{ik}}\right] \,,
\end{align}

\begin{align}\label{eq:nnlo_splitting_kernels_4}
Q^{(nab)}_{ij,k} &= \frac{s_{ijk}^2}{2
   s_{ij} s_{ik}}
   \left[\frac{(1-\epsilon ) z_j^2+2
   \left(1-z_j\right)}{1-z_k}+\frac{(1-\epsilon ) \left(1-z_k\right){}^2+2 z_k}{z_j}\right]
   \\
   \nonumber&
   -\frac{s_{ijk}^2}{4 s_{ik} s_{jk}}
   z_k \left[\frac{(1-\epsilon ) \left(1-z_k\right){}^2+2
   z_k}{z_i z_j}+(1-\epsilon ) \epsilon \right]
   \\
 \nonumber  &
   +\frac{s_{ijk}}{2 s_{ij}}
   \left[ (1-\epsilon ) \frac{ z_i \left(z_i^2-2 z_i+2\right)-z_j \left(z_j^2-6
   z_j+6\right)}{z_j \left(1-z_k\right)}+ 2\epsilon \frac{  z_k \left(z_i-2
   z_j\right)-z_j}{z_j \left(1-z_k\right)}\right]
   \\
\nonumber   &
   +\frac{s_{ijk}}{2 s_{ik}}
   \left[-\epsilon  \left(-z_i+\frac{2 \left(1-z_j\right) \left(z_j-z_k\right)}{z_j
   \left(1-z_k\right)}+z_j\right)-\frac{\left(1-z_i\right) z_k+\left(1-z_j\right){}^3}{z_i
   z_j}
   \right.
   \\
  \nonumber &\;\;\;\;\;\,\,\,\;\;\;\;\;\;
   \left.
   +\epsilon  \left(1-z_j\right) \left(\frac{z_i^2+z_j^2}{z_i z_j}-\epsilon
   \right)+ (1-\epsilon )\frac{ \left(1-z_j\right){}^3-z_j+z_k^2}{z_j
   \left(1-z_k\right)}\right]
   \\
 \nonumber  &+(1-\epsilon ) \left[\frac{t_{ij,k}^2}{4
   s_{ij}{}^2}-\frac{\epsilon }{2}+\frac{1}{4}\right] \,,
\end{align}
where $T_R=\frac{1}{2}$, and  $C_F$ and $C_A$ denote the quadratic Casimirs of the fundamental and adjoint representations of $\textrm{SU}(N_c)$, 
\beq
C_F = \frac{N_c^2-1}{2N_c} \textrm{~~and~~} C_A = N_c\,.
\eeq
The quantity $t_{ij,k}$ is defined as
\beq
t_{ij,k} = 2 \frac{z_is_{jk}-z_j s_{ik}}{z_i+z_j} + \frac{z_i-z_j}{z_i+z_k}s_{ij} \,.
\eeq
Note that the combination,
\begin{equation}
    \frac1{N_c} = C_A - 2C_F\,,
    \label{eq:subleadnc}
\end{equation}
occurs in eq.~(\ref{eq:P23}) as well as later in the text and signals contributions which are sub-leading in $N_c$.

In the case where the parent parton is a gluon, the helicity-tensor is no longer diagonal. In order to simplify the discussion as much as possible, for the rest of the paper we will work in the axial gauge, where the gluon field is subject to the following two conditions:
\beq
\partial_{\mu}A^{\mu} = n_{\mu}A^{\mu} = 0\,,
\eeq
where $n$ is an arbitrary light-like reference vector. In this gauge, the gluon propagator takes the form
\beq\label{eq:axial_propagator}
\frac{i\,\delta^{ab}\,d^{\mu\nu}(p,n)}{p^2+i\varepsilon}\,,\qquad d^{\mu\nu}(p,n) = -g^{\mu\nu}+\frac{p^\mu n^\nu+n^\mu p^{\nu}}{p\cdot n}\,.
\eeq
In principle, we may choose a different reference vector for every gluon (external or internal), as long as the reference vector is not orthogonal to the momentum. In our case, it is convenient to choose all gauge reference vectors to coincide with the reference vector $n$ appearing in the definition of the collinear limit in eq.~\eqref{eq:collinear_momenta}. With this gauge choice, we can write the collinear factorisation for a parent gluon in terms of Lorentz indices rather than helicities~\cite{Catani:1999ss},
\beq
\mathscr{C}_{1\ldots m}\left|\cM_{f_1\ldots f_n}(p_1,\ldots,p_n)\right|^2 = \Bigg(\frac{2\mu^{2\epsilon}\,g_s^2}{s_{1\ldots m}}\Bigg)^{m-1} \,\hat{P}^{\mu\nu}_{f_1\ldots f_m}\, \mathcal{T}_{ff_{m+1}\ldots f_n,\mu\nu}(\pM{},p_{m+1},\ldots,p_n)\,,
\label{eq:gluon_factorisation}
\eeq
where quantities with open Lorentz indices are obtained by amputating the polarisation vectors. 
The formulation in terms of Lorentz indices has the advantage that we do not need to work with the extra-dimensional physical polarisation states of the gluon.
The complete tensor structure of the splitting amplitude contains terms involving the transverse momenta of the collinear partons~\cite{Catani:1999ss},
\beq\label{eq:P_tensor}
\hat{P}^{\mu\nu}_{f_1\ldots f_m} = g^{\mu\nu}\,A^{(g)}_{f_1\ldots f_m} + \sum_{i,j=1}^m\,\frac{\tilde{k}_{\perp i}^{\mu}\tilde{k}_{\perp j}^{\nu}}{s_{1\ldots m}}\,B^{(g)}_{ij,f_1\ldots f_m}\,.
\eeq
Let us make a comment about gauge invariance. Since physical polarisation states are transverse, only the transverse part of a Lorentz tensor carries physical information (because the non-transverse part vanishes upon contraction with a physical polarisation vector). For this reason it is often sufficient to consider Lorentz tensors that are explicitly transverse. This is the case of the tensor $\mathcal{T}_{ff_{m+1}\ldots f_n}^{\mu\nu}$ in eq.~\eqref{eq:gluon_factorisation}, which can be chosen to be explicitly transverse, i.e., it can be chosen to satisfy
\beq
\pM{}_\mu\,\mathcal{T}_{ff_{m+1}\ldots f_n}^{\mu\nu} = \pM{}_\nu\,\mathcal{T}_{ff_{m+1}\ldots f_n}^{\mu\nu} = 0\,.
\eeq
The splitting amplitude in eq.~\eqref{eq:P_tensor}, however, is in general not transverse, because we have
\beq
\pM{}_{\mu}\,\hat{P}^{\mu\nu}_{f_1\ldots f_m} = \pM{\nu}\,A^{(g)}_{f_1\ldots f_m}\,.
\eeq
This is immaterial so long as the splitting amplitude is contracted to a tensor $\mathcal{T}_{ff_{m+1}\ldots f_n}^{\mu\nu}$ that is transverse. We could alternatively have defined an explicitly transverse splitting amplitude by replacing $g^{\mu\nu}$ by $-d^{\mu\nu}(\pM{},n)$ in eq.~\eqref{eq:P_tensor}, at the price of introducing terms proportional to the gauge vector $n$ that cancel when contracted with a transverse quantity.

\section{Quark-parent splitting amplitudes for four collinear partons}
\label{sec:collinear}

In this section we present the main result of our paper, namely the computation of the tree-level splitting amplitudes for $m=4$ collinear partons,
in the case where the parent parton is a quark. The computation follows the same lines as the computation of the case $m=3$ in ref.~\cite{Catani:1999ss}, and we review the different steps for completeness in the remainder of this section. Our results for the splitting amplitudes are rather lengthy. We therefore do not present all of them here in printed form, but we make them available in computer-readable form~\cite{QuadColKernelsWebsite}.
Note that the constraints in eq.~(\ref{eq:constraint})
have not been imposed on the splitting amplitudes.
This may allow us, through crossing symmetry, to readily 
obtain the splitting amplitudes for initial-state collinear emissions~\cite{deFlorian:2001zd}. 

{
}

{
}

Let us start by describing the steps we perform to compute the splitting amplitudes. Our goal is to isolate the leading divergent behaviour of a tree-level amplitude in the limit where $m$ partons become collinear, defined as the limit where their transverse momenta $k_{\perp i}$ in eq.~\eqref{eq:collinear_momenta} approach zero at the same rate. In order to isolate the leading behaviour, we introduce a small parameter $\lambda$ and perform the uniform rescaling $k_{\perp i} \to \lambda k_{\perp i}$, $1\le i\le m$. After this rescaling the matrix element depends on $\lambda$, and we can approach the collinear limit by expanding the matrix element into a Laurent series around $\lambda=0$. The leading term in the expansion, which corresponds to the coefficient of $1/\lambda^{2(m-1)}$, is universal and described by the collinear factorisation in eq.~\eqref{eq:factorisation}. 

In applying the previous algorithm, we could start from any amplitude for $n\ge m+3$ partons, work out the interferences between all Feynman diagrams in $D$ dimensions and only keep their contribution to the leading term in $\lambda$. In ref.~\cite{Catani:1999ss} it was argued that when working in a physical gauge (e.g., axial gauge) only a subset of diagrams needs to be considered. Indeed, in a physical gauge contributions from Feynman diagrams where collinear partons are separated by a hard propagator are subleading in the collinear limit. As a consequence, we only need to consider a subset of Feynman diagrams in the axial gauge, and we have
\beq\bsp
\mathscr{C}_{1\ldots m}&\left|\cM_{f_1\ldots f_n}(p_1,\ldots,p_n)\right|^2 = \\
&\mathscr{C}_{1\ldots m}\left[
 \Bigg(\frac{2\mu^{2\epsilon}g_s^2}{s_{1\ldots m}}\Bigg)^{m-1} 
\left[ \mathcal{M}^{(n)\, s}_{ff_{m+1}\ldots f_n}\right]^\ast V^{(n)\, ss'}_{f_1\ldots f_m}(p_1,\ldots,p_m) \mathcal{M}^{(n)\, s'}_{ff_{m+1}\ldots f_n}\right]\,,
\label{eq:V_fac2}
\esp\eeq
where a sum over the spin indices $s$, $s'$ of the intermediate state is understood, and we suppress all colour and spin indices of the external partons. Here $\mathcal{M}^{(n)\, s}_{ff_{m+1}\ldots f_n} = \mathcal{M}^{(n)\, s}_{ff_{m+1}\ldots f_n}(P,p_{m+1},\ldots,p_n)$ denotes the sum of all Feynman diagrams with an off-shell leg with momentum $P$, flavour $f$ and spin $s$. Note that this subset of Feynman diagrams is by itself not gauge invariant, and the superscript $(n)$ indicates the dependence on the gauge choice. 
The squared off-shell current $ V^{(n)\, ss'}_{f_1\ldots f_m}$ may be written as the interference of two colour-dressed off-shell currents,
\begin{equation}\label{eq:V_to_j}
\!\!\Bigg(\frac{2\mu^{2\epsilon}g_s^2}{s_{1\ldots m}}\Bigg)^{m-1} V^{(n)\, ss'}_{f_1\ldots f_m}(p_1,\ldots,p_m) =
\frac1{\mathcal{C}_{f}} \sum_{\substack{(s_1,\ldots,s_m)\\(c,c_1,\ldots,c_m)}}\,
\left[{\mathrm J}_{f_1\ldots f_m}^{c,c_1\ldots c_m;s's_1\ldots s_m}\right]^{\ast}\,
{\mathrm J}_{f_1\ldots f_m}^{c,c_1\ldots c_m;ss_1\ldots s_m}\,,
\end{equation}
where $\mathcal{C}_{f}$ is defined after eq.~(\ref{eq:P_def}). Note that also $ V^{(n)\, ss'}_{f_1\ldots f_m}$ depends on the gauge vector $n$ before the collinear limit is taken. Since the collinear limit is gauge invariant, this dependence disappears in the limit, and the squared off-shell current reduces to the splitting amplitude,
\beq\label{eq:V_to_P}
\mathscr{C}_{1\ldots m}V^{(n)\, ss'}_{f_1\ldots f_m}(p_1,\ldots,p_m) = \hat{P}_{f_1\ldots f_m}^{ss'}\,.
\eeq

Using eq.~\eqref{eq:V_to_P} we can substantially reduce the number of interfering Feynman diagrams that we need to evaluate. This strategy can always be used, independently of the flavour of the parent parton. We focus in this paper solely on the case where the parent parton is a quark, $f=q$. In that situation we can further simplify the computation by averaging over the spins of the parent quark and use eq.~\eqref{eq:unpol} to recover the polarised splitting amplitude from the unpolarised one. Averaging over fermion spins is equivalent to computing the trace of Dirac spin indices, and we obtain the following simple formula relating the unpolarised splitting amplitude to the squared off-shell current,
\begin{equation}
\langle \hat{P}_{f_1\ldots f_m} \rangle = \mathscr{C}_{1\ldots m}\left[\textrm{Tr}\Bigg(\frac{\slashed{n} V^{(n)}_{f_1\ldots f_m}(p_1,\ldots,p_m)}{4 P \cdot n}\Bigg)\right]\,.
\label{eq:CG42}
\end{equation}
We have used this procedure to compute all quark-initiated splitting amplitudes up to $m=4$, and we reproduce all known results for the cases $m=2$ and 3 in eq.~\eqref{eq:P23}. The results for $m=4$ are new and are presented for the first time in this paper. More precisely, there are three different quark-initiated splitting amplitudes of the form $q\to f_1f_2f_3q$, with $(f_1,f_2,f_3)\in\{(\bar{q}',q',g),(\bar{q},q,g),(g,g,g)\}$. Splitting amplitudes for anti-quark initiated processes can easily be obtained from charge conjugation. In the remainder of this section we discuss in more detail the computation of these splitting amplitudes. The explicit results are available in computer-readable form~\cite{QuadColKernelsWebsite}.

\begin{figure}[!t]
\centering
\begin{align*}
&
\begin{tikzpicture}[baseline=(a)]
\begin{feynman}[inline=(a)]
\node[blob, scale=1] (blob1);
\vertex [right=0.3cm of blob1] (a);
\vertex [right=0.8cm of a] (b);
\vertex [above right= 1cm of b] (g1);
\vertex [above right= 1cm of g1] (q1) {\(\bar{q}^{\prime}_1\)};
\vertex [right= 1cm of g1] (q2) {\(q^{\prime}_2\)};
\vertex [right=2cm of b] (c);
\vertex [above right= 2cm of c] (g2) {\(g_3\)};
\vertex [right=1.5cm of c] (d) {\(q_4\)};
\diagram*{
(blob1)-- (a) -- [fermion] (b) -- [gluon] (g1),
(g1) -- [anti fermion] (q1),
(g1) -- [fermion] (q2),
(b) -- [fermion] (c) -- [gluon] (g2),
(c) -- [fermion] (d)
};
\end{feynman}
\end{tikzpicture} 
\begin{tikzpicture}[baseline=(a)]
\begin{feynman}[inline=(a)]
\node[blob, scale=1] (blob1);
\vertex [right=0.3cm of blob1] (a);
\vertex [left=1cm of blob1] (pl){\(+\)};
\vertex [right=0.8cm of a] (b);
\vertex [above right= 2cm of b] (g2) {\(g_3\)};
\vertex [right=1.5cm of b] (c);
\vertex [above right= 1cm of c] (g1);
\vertex [above right= 1cm of g1] (q1) {\(\bar{q}^{\prime}_1\)};
\vertex [right= 1cm of g1] (q2) {\(q^{\prime}_2\)};
\vertex [right=1.5cm of c] (d) {\(q_4\)};
\diagram*{
(blob1)-- (a) -- [fermion] (b) -- [gluon] (g2),
(b) -- [fermion] (c) -- [gluon] (g1),
(g1) -- [anti fermion] (q1),
(g1) -- [fermion] (q2),
(c) -- [fermion] (d)
};
\end{feynman}
\end{tikzpicture} \\
&
\begin{tikzpicture}[baseline=(a)]
\begin{feynman}[inline=(a)]
\node[blob, scale=1] (blob1);
\vertex [left=1cm of blob1] (pl){\(+\)};
\vertex [right=0.3cm of blob1] (a);
\vertex [right=0.8cm of a] (b);
\vertex [above right= 0.75cm of b] (g11);
\vertex [above right= 0.75cm of g11] (g12);
\vertex [above right= 1cm of g12] (q1) {\(\bar{q}^{\prime}_1\)};
\vertex [right=1.5cm of g11] (g3) {\(g_3\)};
\vertex [right= 1cm of g12] (q2) {\(q^{\prime}_2\)};
\vertex [right=2cm of b] (c) {\(q_4\)};
\diagram*{
(blob1)-- (a) -- [fermion] (b) --  [gluon] (g12),
(g12) -- [anti fermion] (q1),
(g12) -- [fermion] (q2),
(g11)-- [gluon] (g3),
(b) -- [fermion] (c)
};
\end{feynman}
\end{tikzpicture} 
\begin{tikzpicture}[baseline=(a)]
\begin{feynman}[inline=(a)]
\node[blob, scale=1] (blob1);
\vertex [right=0.3cm of blob1] (a);
\vertex [left=1cm of blob1] (pl){\(+\)};
\vertex [right=0.8cm of a] (b);
\vertex [above right= 1cm of b] (g1);
\vertex [above right= 0.5cm of g1] (q11);
\vertex [above right= 1cm of q11] (q12) {\(\bar{q}^{\prime}_1\)};
\vertex [right= 1cm of q11] (g3) {\(g_3\)};
\vertex [right= 1.5cm of g1] (q2) {\(q^{\prime}_2\)};
\vertex [right=2cm of b] (c) {\(q_4\)};
\diagram*{
(blob1)-- (a) -- [fermion] (b) -- [gluon] (g1),
(g1) -- (q11) -- [anti fermion] (q12),
(q11) -- [gluon] (g3),
(g1) -- [fermion] (q2),
(b) -- [fermion] (c)
};
\end{feynman}
\end{tikzpicture} 
\begin{tikzpicture}[baseline=(a)]
\begin{feynman}[inline=(a)]
\node[blob, scale=1] (blob1);
\vertex [left=1cm of blob1] (pl){\(+\)};
\vertex [right=0.3cm of blob1] (a);
\vertex [right=0.8cm of a] (b);
\vertex [above right= 1cm of b] (g1);
\vertex [above right= 1.5cm of g1] (q1) {\(\bar{q}^{\prime}_1\)};
\vertex [right= 0.5cm of g1] (q21);
\vertex [above right= 1cm of q21] (g3) {\(g_3\)};
\vertex [right= 1cm of q21] (q22) {\(q^{\prime}_2\)};
\vertex [right=2cm of b] (c) {\(q_4\)};
\diagram*{
(blob1)-- (a) -- [fermion] (b) -- [gluon] (g1),
(g1) -- [anti fermion] (q1),
(g1) -- (q21) -- [fermion] (q22),
(q21) --  [gluon] (g3),
(b) -- [fermion] (c)
};
\end{feynman}
\end{tikzpicture} 
\end{align*} 
\caption{The Feynman diagrams contributing to the off-shell current $q \to \bar{q}^\prime_1 q^\prime_2  g_3  q_4$. In the case of identical quarks, $q'_2=q_4$, we also need to include the diagrams with $q'_2$ and $q_4$ exchanged. }
\label{fig:qqgqPrime}
\end{figure}
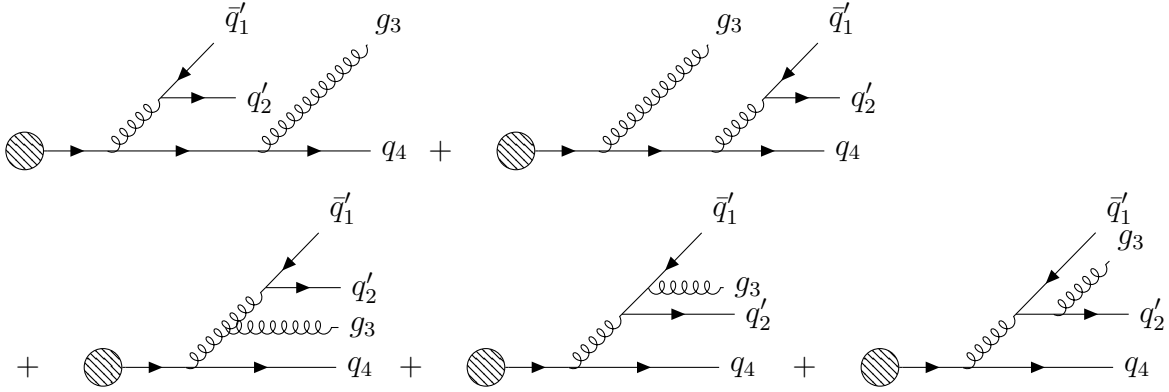

Let us start by discussing the collinear splitting $q \to \bar{q}' q' g q$. There are five diagrams that contribute to the off-shell current ${\mathrm J}_{\bar{q}' q' g q}$ in eq.~\eqref{eq:V_to_j}. The diagrams are shown in fig. \ref{fig:qqgqPrime}.
Going through the steps outlined above, we find that the result for the splitting amplitude $q \to \bar{q}' q' g q$
can be decomposed into an `abelian' and a `non-abelian' part,
\begin{equation}
\label{eq:Pq_qb'q'gq}
\langle \hat{P}_{\bar{q}^\prime_1q^\prime_2g_3q_4} \rangle = \frac{1}{2}C_F^2\langle \hat{P}^{(\text{ab})}_{\bar{q}^\prime_1q^\prime_2g_3q_4} \rangle+ \frac{1}{2}C_AC_F\langle \hat{P}^{(\text{nab})}_{\bar{q}^\prime_1q^\prime_2g_3q_4}\rangle\,.
\end{equation}
The indices carried by the parton label refer to the indices of the momenta and the momentum fractions of the partons.

In the case where the quarks in the final state have the same flavour, $q'=q$, we need to include also Feynman diagrams where the quarks 2 and 4 in fig.~\ref{fig:qqgqPrime} are interchanged. This naturally leads to the following representation of the splitting amplitude $q \to \bar{q} q g q$,
\begin{equation}
\langle \hat{P}_{\bar{q}_1q_2g_3q_4}\rangle = \Big[\langle \hat{P}_{\bar{q}^\prime_1q_2^\prime g_3q_4}\rangle + (2 \leftrightarrow 4)\Big] + \langle \hat{P}_{\bar{q}_1q_2g_3q_4}^{(\text{id})}\rangle\,.
\label{eq:Pqqgq}
\end{equation}
The terms in square brackets denote the contributions from the splitting amplitude for different quark flavours in eq.~\eqref{eq:Pq_qb'q'gq}. The last term is new, and captures interference contributions from identical quarks. We can decompose it into contributions with different colour factors, corresponding again to `abelian' and `non-abelian' parts,
\begin{equation}
\langle \hat{P}_{\bar{q}_1q_2g_3q_4}^{(\text{id})}\rangle = \frac{1}{2}C_F^2(2\,C_F-C_A)\langle \hat{P}_{\bar{q}_1q_2g_3q_4}^{(\text{id})(\text{ab})}\rangle + \frac{1}{4}C_AC_F(C_A-2C_F)\langle \hat{P}_{\bar{q}_1q_2g_3q_4}^{(\text{id})(\text{nab})}\rangle.
\label{eq:P_q1q2g3q4(id)}
\end{equation}
Note that, because of eq.~(\ref{eq:subleadnc}), the two terms on the right-hand side of eq.~(\ref{eq:P_q1q2g3q4(id)}) yield a contribution to eq.~(\ref{eq:Pqqgq}) which is sub-leading in $N_c$.

\begin{figure}[!t]
\centering
\begin{equation} \nonumber
\begin{split} 
&
\begin{tikzpicture}[baseline=(a)]
\begin{feynman}[inline=(a)]
\node[blob, scale=1] (blob1);
\vertex [left=0.5cm of blob1] (bracket){\(\Big\{\)};
\vertex [right=0.3cm of blob1] (a);
\vertex [right=0.7cm of a] (b);
\vertex [above right= 2cm of b] (g1) {\(g_1\)};
\vertex [right=1cm of b] (c);
\vertex [above right= 2cm of c] (g2) {\(g_2\)};
\vertex [right=1cm of c] (d);
\vertex [above right= 2cm of d] (g3) {\(g_3\)};
\vertex [right=1.5cm of d] (e) {\(q_4\)};
\vertex [right=2cm of d] (perm){\(+ \text{ (5 permutations) }\)};
\vertex [right=2.1cm of perm] (bracket){\(\Big\} \text{ +}\)};
\diagram*{
(blob1)-- (a) -- [fermion] (b) -- [gluon] (g1),
(b) -- [fermion] (c) -- [gluon] (g2),
(c) -- [fermion] (d) -- [gluon] (g3),
(d) -- [fermion] (e)
};
\end{feynman}
\end{tikzpicture}  
\begin{tikzpicture}[baseline=(a)]
\begin{feynman}[inline=(a)]
\node[blob, scale=1] (blob1);
\vertex [right=0.3cm of blob1] (a);
\vertex [right=1.5cm of a] (b);
\vertex [above right=1cm of b] (c);
\vertex [below right=1.5cm of b] (d) {\(q_4\)};
\vertex [above right=1cm of c] (g1){\(g_1\)};
\vertex [right=1cm of c] (g2){\(g_2\)};
\vertex [below right=1cm of c] (g3){\(g_3\)};
\diagram*{
(blob1) -- (a) -- [fermion] (b) -- [fermion] (d),
(b) -- [gluon] (c) -- [gluon] (g1),
(c) -- [gluon] (g2),
(c) -- [gluon] (g3)
};
\end{feynman}
\end{tikzpicture} \\
&
\begin{tikzpicture}[baseline=(a)]
\begin{feynman}[inline=(a)]
\node[blob, scale=1] (blob1);
\vertex [left=0.5cm of blob1] (bracket){\(\Big\{\)};
\vertex [left=1cm of blob1] (pl){\(+\)};
\vertex [right=0.3cm of blob1] (a);
\vertex [right=0.7cm of a] (b);
\vertex [above right= 1cm of b] (g11);
\vertex [above right= 1cm of g11] (g12){\(g_1\)};
\vertex [right= 0.7cm of g11] (g2) {\(g_2\)};
\vertex [right=1.5cm of b] (c);
\vertex [above right= 2cm of c] (g3) {\(g_3\)};
\vertex [right=1.5cm of c] (d) {\(q_4\)};
\vertex [right=0.5cm of d] (pr){\(+\)};
\diagram*{
(blob1)-- (a) -- [fermion] (b) -- [gluon] (g11),
(g11)-- [gluon] (g12),
(g11) -- [gluon] (g2),
(b) -- [fermion] (c) -- [gluon] (g3),
(c) -- [fermion] (d)
};
\end{feynman}
\end{tikzpicture} 
\begin{tikzpicture}[baseline=(a)]
\begin{feynman}[inline=(a)]
\node[blob, scale=1] (blob1);
\vertex [right=0.3cm of blob1] (a);
\vertex [right=0.7cm of a] (b);
\vertex [above right= 2cm of b] (g3) {\(g_3\)};
\vertex [right=1cm of b] (c);
\vertex [above right= 1cm of c] (g);
\vertex [above right= 1cm of g] (g1){\(g_1\)};
\vertex [right= 0.7cm of g] (g2) {\(g_2\)};
\vertex [right=1.5cm of c] (d) {\(q_4\)};
\vertex [right=2.2cm of d] (perm){\(+\text{ }(1 \leftrightarrow 3)+(2 \leftrightarrow 3)\)};
\vertex [right=2.1cm of perm] (bracket){\(\Big\}\)};
\diagram*{
(blob1)-- (a) -- [fermion] (b) -- [gluon] (g3),
(b) -- [fermion] (c) -- [gluon] (g),
(g)-- [gluon] (g1),
(g) -- [gluon] (g2),
(c) -- [fermion] (d)
};
\end{feynman}
\end{tikzpicture} \\
&
\begin{tikzpicture}[baseline=(a)]
\begin{feynman}[inline=(a)]
\node[blob, scale=1] (blob1);
\vertex [right=0.3cm of blob1] (a);
\vertex [left=1cm of blob1] (pl){\(+\)};
\vertex [left=0.5cm of blob1] (bracket){\(\Big\{\)};
\vertex [right=1.5cm of a] (b);
\vertex [above right=1.2cm of b] (c);
\vertex [below right=1.5cm of b] (d) {\(q_4\)};
\vertex [above right=1cm of c] (g1){\(g_1\)};
\vertex [right=0.8cm of c] (g2){\(g_2\)};
\vertex [above right=0.5cm of b] (e);
\vertex [right=1.3cm of e] (g3){\(g_3\)};
\vertex [right=2.3cm of b] (perm){\(+\text{ }(1 \leftrightarrow 3)+(2 \leftrightarrow 3)\)};
\vertex [right=2.1cm of perm] (bracket){\(\Big\}\)};
\diagram*{
(blob1) -- (a) -- [fermion] (b) -- [fermion] (d),
(b) -- [gluon] (g1),
(c) -- [gluon] (g2),
(e) -- [gluon] (g3)
};
\end{feynman}
\end{tikzpicture}
\end{split}
\end{equation}
\caption{The Feynman diagrams contributing to the off-shell current $q\to g_1g_2g_3q_4$.}
\label{fig:gggq}
\end{figure}

Finally let us discuss the splitting amplitude $q \to g g g q$. The Feynman diagrams contributing to the off-shell current are shown in fig.~\ref{fig:gggq}. We can decompose the splitting amplitude into contributions from different colour factors as follows,
\begin{equation}
\begin{split}
\langle \hat{P}_{g_1g_2g_3q_4} \rangle = C_F^3 \langle \hat{P}_{g_1g_2g_3q_4}^{(\text{ab})} \rangle + C_F^2C_A \langle \hat{P}_{g_1g_2g_3q_4}^{(\text{nab})_1} \rangle + \frac{3}{2}C_A^2C_F \langle \hat{P}_{g_1g_2g_3q_4}^{(\text{nab})_2} \rangle.
\end{split}
\label{eq:colour_structurePgggq}
\end{equation}
We note here that the computation of $\hat{P}_{gggq}$ poses a challenge due to the large number of interference diagrams that need to be evaluated once all gluon permutations are taken into account. An important step in the computation was therefore to exploit as much as possible symmetries between the different permutations in order to minimise the number of terms.
In particular, one may exploit the symmetries under the exchange of the three external gluons to reduce the number of diagrams that need to be directly evaluated. We can thus write $\hat{P}_{gggq}$ in a symmetrised form as

\begin{equation}
\langle \hat{P}_{g_1g_2g_3q_4} \rangle =
\langle \hat{P}_{g_1g_2g_3q_4}^{\text{ symm.}} \rangle
+ (5 \text{ permutations of } g_1g_2g_3 )\,,
\end{equation}
which can be decomposed into different colour factors as in eq.~\eqref{eq:colour_structurePgggq},
\begin{equation}
\begin{split}
\langle \hat{P}_{g_1g_2g_3q_4}^{\text{ symm.}} \rangle = 
C_F^3 \langle \hat{P}_{g_1g_2g_3q_4}^{(\text{ab})\text{ symm.}} \rangle 
+ C_F^2C_A \langle \hat{P}_{g_1g_2g_3q_4}^{(\text{nab})_1\text{ symm.}} \rangle 
+ \frac{3}{2}C_A^2C_F \langle \hat{P}_{g_1g_2g_3q_4}^{(\text{nab})_2\text{ symm.}} \rangle.
\end{split}
\label{eq:colour_structurePgggqsymm}
\end{equation}

{
}


{
}

{
}



\section{Nested collinear limits}
\label{sec:nested_collinear}
\subsection{Strongly-ordered vs. iterated collinear limits}

In this section we analyse the collinear limit of the splitting amplitudes themselves, i.e., we study their behaviour in the limit where a subset of collinear partons is more collinear than the others. Since splitting amplitudes are gauge-invariant quantities that share many properties with on-shell scattering amplitudes, they must themselves exhibit a factorisation similar to eq.~\eqref{eq:factorisation} for amplitudes. The reasons for studying these limits are twofold: Firstly, checking that our splitting amplitudes have the correct properties under collinear limits is a strong sanity check on our results. Secondly, the knowledge of these limits is an important ingredient whenever splitting amplitudes are used to build counterterms to subtract infrared divergences in higher-order computations. 

\begin{figure}[!t]
\centering
\begin{equation} \nonumber
\begin{tikzpicture}[baseline=(a),scale=0.9, every node/.style={transform shape}]
\begin{feynman}[inline=(a)]
\node[circle,scale=1,fill=black!20,draw=black,thick] (dot) {\(\mathcal{M}\)};
\vertex [right=0.2cm of dot] (a);
\vertex [above = 0.3cm of a, draw=none] (11null);
\vertex [below = 0.3cm of a, draw=none] (21null);
\vertex [right = 0.2cm of 11null, draw=none] (12null);
\vertex [right = 0.2cm of 21null, draw=none] (22null);
\vertex [right = 1.2cm of 12null] (c11);
\vertex [above = 0.5cm of c11] (c12) {\(p_1\)};
\vertex [ right = 0.8cm of a] (cdot) {\(\cdot\)};
\vertex [ right = 2cm of a] (to) {\(\to\)};
\vertex [right = 1.2cm of 22null] (c21);
\vertex [below = 0.5cm of c21] (c22) {\(p_n\)};
\diagram*{
(12null) -- [edge label' = \(\cdot\)] (c12),
(22null) -- [edge label = \(\cdot\)] (c22)
};
\end{feynman}
\end{tikzpicture}
\quad
\begin{tikzpicture}[baseline=(a),scale=0.9, every node/.style={transform shape}]
\begin{feynman}[inline=(a)]
\node[circle,scale=1,fill=black!20,draw=black,thick] (dot) {\(\mathcal{M}\)};
\vertex [below right = 0.5cm of dot, draw=none] (dummy1);
\vertex [below left = 0.2cm of dummy1, draw=none] (dummy2);
\vertex [below = 0.5cm of dot, draw=none] (dummy3);
\vertex [below right= 0.4cm of dummy1] (pmplus1) {\(p_{m+1}\)};
\vertex [below = 0.5cm of dummy2] (pn) {\(p_{n}\)};
\vertex [right=of dot] (a);
\vertex [right = 0.75cm of a] (b);
\fill[black!70] (2.2,0) ellipse (0.5 and 1.5);
\node[circle, scale=2, right = 0.9cm of a,draw=none] (b2);
\vertex [above = 1cm of b2, draw=none] (11null);
\vertex [right = 0.25cm of 11null, draw=none] (12null);
\vertex [below = 1cm of b2, draw=none] (21null);
\vertex [right = 0.25cm of 21null, draw=none] (22null);
\vertex [right = 1.5cm of 12null] (c1) {\(p_1\)};
\vertex [ right = 1.2cm of b2] (cdot) {\(\cdot\)};
\vertex [ right = 3cm of b2] (to) {\(\to\)};
\vertex [right = 1.5cm of 22null] (c2) {\(p_m\)};
\diagram*{
(dummy1) -- [edge label' = \(\cdot\)] (pmplus1), 
(dummy3) -- (pn), 
(dot) -- (a) -- [edge label=\(\pM{}\)] (b),
(12null) -- [edge label' = \(\cdot\)] (c1),
(22null) -- [edge label = \(\cdot\)] (c2)
};
\end{feynman}
\end{tikzpicture}
\quad
\begin{tikzpicture}[baseline=(a),scale=0.9, every node/.style={transform shape}]
\begin{feynman}[inline=(a)]
\node[circle,scale=1,fill=black!20,draw=black,thick] (dot) {\(\mathcal{M}\)};
\vertex [below right = 0.5cm of dot, draw=none] (dummy1);
\vertex [below left = 0.2cm of dummy1, draw=none] (dummy2);
\vertex [below = 0.5cm of dot, draw=none] (dummy3);
\vertex [below right= 0.4cm of dummy1] (pmplus1) {\(p_{m+1}\)};
\vertex [below = 0.5cm of dummy2] (pn) {\(p_{n}\)};
\vertex [right=of dot] (a);
\vertex [right = 0.75cm of a] (b);
\fill[black!70] (2.2,0) ellipse (0.5 and 1.5);
\node[circle, scale=2, right = 0.9cm of a,draw=none] (b2);
\vertex [above = 1cm of b2, draw=none] (11null);
\vertex [right = 0.25cm of 11null, draw=none] (12null);
\vertex [below = 1cm of b2, draw=none] (21null);
\vertex [right = 0.25cm of 21null, draw=none] (22null);
\vertex [right = 1.5cm of 12null] (c1);
\vertex [right = 1.7cm of 12null] (split);
\vertex [ right = 1.2cm of b2] (cdot) {\(\cdot\)};
\vertex [right = 1.5cm of 22null] (c2) {\(p_m\)};
\fill[black!70] (split) ellipse (0.2 and 1);
\vertex [above =0.5cm of split,draw=none] (dummyf1);
\vertex [right = 0.15cm of dummyf1,draw=none] (of1);
\vertex [right = 0.5cm of of1] (f1) {\(p_1\)};
\vertex [below =0.5cm of split,draw=none] (dummyfn);
\vertex [right = 0.15cm of dummyfn,draw=none] (ofn);
\vertex [right = 0.5cm of ofn] (fn) {\(p_{m'}\)};
\vertex [right = 0.3cm of split] (cdot2) {\(\cdot\)};
\diagram*{
(dummy1) -- [edge label' = \(\cdot\)] (pmplus1), 
(dummy3) -- (pn), 
(dot) -- (a) -- [edge label=\(\pM{}\)] (b),
(12null) -- [edge label' = \(\cdot\),edge label= \(\pN{}\)] (c1),
(22null) -- [edge label = \(\cdot\)] (c2),
(of1) -- [edge label' = \(\cdot\)] (f1),
(ofn) -- [edge label = \(\cdot\)](fn)
};
\end{feynman}
\end{tikzpicture}
\end{equation}
\caption{Consecutive splitting $f \to f_1+ \ldots + f_{m'} + \ldots + f_{m} \to (f_1 +\ldots +f_{m'}) +\ldots + f_m$ where an $m'$-parton subset becomes collinear to the direction $\pN{}$ within the larger $m$-parton collinear set.}
\label{fig:nested_split1}
\end{figure}
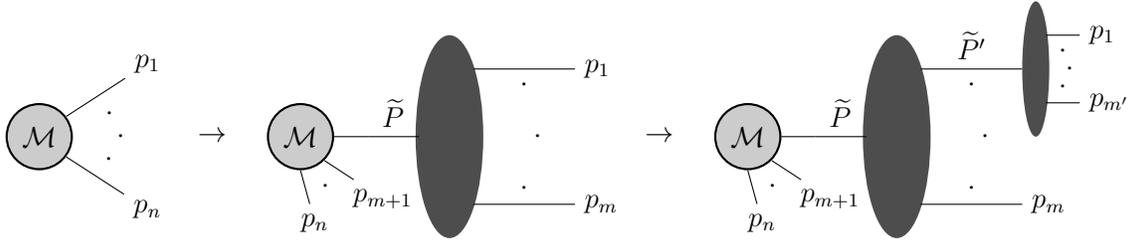

\begin{figure}[H]
\centering
\begin{equation} \nonumber
\begin{tikzpicture}[baseline=(a),scale=0.9, every node/.style={transform shape}]
\begin{feynman}[inline=(a)]
\node[circle,scale=1,fill=black!20,draw=black,thick] (dot) {\(\mathcal{M}\)};
\vertex [right=0.2cm of dot] (a);
\vertex [above = 0.3cm of a, draw=none] (11null);
\vertex [below = 0.3cm of a, draw=none] (21null);
\vertex [right = 0.2cm of 11null, draw=none] (12null);
\vertex [right = 0.2cm of 21null, draw=none] (22null);
\vertex [right = 1.2cm of 12null] (c11);
\vertex [above = 0.5cm of c11] (c12) {\(p_1\)};
\vertex [ right = 0.8cm of a] (cdot) {\(\cdot\)};
\vertex [ right = 2cm of a] (to) {\(\to\)};
\vertex [right = 1.2cm of 22null] (c21);
\vertex [below = 0.5cm of c21] (c22) {\(p_n\)};
\diagram*{
(12null) -- [edge label' = \(\cdot\)] (c12),
(22null) -- [edge label = \(\cdot\)] (c22)
};
\end{feynman}
\end{tikzpicture}
\quad
\begin{tikzpicture}[baseline=(a),scale=0.9, every node/.style={transform shape}]
\begin{feynman}[inline=(a)]
\node[circle,scale=1,fill=black!20,draw=black,thick] (dot) {\(\mathcal{M}\)};
\vertex [below right = 0.5cm of dot, draw=none] (dummy1);
\vertex [below left = 0.2cm of dummy1, draw=none] (dummy2);
\vertex [below = 0.5cm of dot, draw=none] (dummy3);
\vertex [below left = 0.5cm of dot, draw=none] (dummy4);
\vertex [below right= 0.4cm of dummy1] (primeplus1) {\(p_{m'+1}\)};
\vertex [below = 0.5cm of dummy2] (pm) {\(p_{m}\)};
\vertex [below left= 0.5cm of dummy4] (pn) {\(p_{n}\)};
\vertex [right=of dot] (a);
\vertex [right = 1cm of a] (b);
\fill[black!70] (2.2,0) ellipse (0.2 and 1);
\node[circle, scale=2, right = 0.9cm of a,draw=none] (b2);
\vertex [above = 0.7cm of b2, draw=none] (11null);
\vertex [below = 0.7cm of b2, draw=none] (21null);
\vertex [right = 0.05cm of 11null, draw=none] (12null);
\vertex [right = 0.05cm of 21null, draw=none] (22null);
\vertex [right = 1.2cm of 12null] (c1) {\(p_1\)};
\vertex [ right = 1cm of b2] (cdot) {\(\cdot\)};
\vertex [ right = 2.6cm of b2] (to) {\(\to\)};
\vertex [right = 1.2cm of 22null] (c2) {\(p_{m'}\)};
\diagram*{
(dummy1) -- [edge label' = \(\cdot\)] (primeplus1), 
(dummy3) -- (pm), 
(dummy4) -- [edge label = \(\cdot\)] (pn), 
(dot) -- (a) -- [edge label=\(\pN{}\)] (b),
(12null) -- [edge label' = \(\cdot\)] (c1),
(22null) -- [edge label = \(\cdot\)] (c2)
};
\end{feynman}
\end{tikzpicture}
\quad
\begin{tikzpicture}[baseline=(a),scale=0.9, every node/.style={transform shape}]
\begin{feynman}[inline=(a)]
\node[circle,scale=1,fill=black!20,draw=black,thick] (dot) {\(\mathcal{M}\)};
\vertex [below right = 0.5cm of dot, draw=none] (dummy1);
\vertex [below left = 0.2cm of dummy1, draw=none] (dummy2);
\vertex [below = 0.5cm of dot, draw=none] (dummy3);
\vertex [below right= 0.4cm of dummy1] (pmplus1) {\(p_{m+1}\)};
\vertex [below = 0.5cm of dummy2] (pn) {\(p_{n}\)};
\vertex [right=of dot] (a);
\vertex [right = 0.75cm of a] (b);
\fill[black!70] (2.2,0) ellipse (0.5 and 1.5);
\node[circle, scale=2, right = 0.9cm of a,draw=none] (b2);
\vertex [above = 1cm of b2, draw=none] (11null);
\vertex [right = 0.25cm of 11null, draw=none] (12null);
\vertex [below = 1cm of b2, draw=none] (21null);
\vertex [right = 0.25cm of 21null, draw=none] (22null);
\vertex [right = 1.5cm of 12null] (c1);
\vertex [right = 1.7cm of 12null] (split);
\vertex [ right = 1.2cm of b2] (cdot) {\(\cdot\)};
\vertex [right = 1.5cm of 22null] (c2) {\(p_m\)};
\fill[black!70] (split) ellipse (0.2 and 1);
\vertex [above =0.5cm of split,draw=none] (dummyf1);
\vertex [right = 0.15cm of dummyf1,draw=none] (of1);
\vertex [right = 0.5cm of of1] (f1) {\(p_1\)};
\vertex [below =0.5cm of split,draw=none] (dummyfn);
\vertex [right = 0.15cm of dummyfn,draw=none] (ofn);
\vertex [right = 0.5cm of ofn] (fn) {\(p_{m'}\)};
\vertex [right = 0.3cm of split] (cdot2) {\(\cdot\)};
\diagram*{
(dummy1) -- [edge label' = \(\cdot\)] (pmplus1), 
(dummy3) -- (pn), 
(dot) -- (a) -- [edge label=\(\pM{}\)] (b),
(12null) -- [edge label' = \(\cdot\),edge label= \(\pN{}\)] (c1),
(22null) -- [edge label = \(\cdot\)] (c2),
(of1) -- [edge label' = \(\cdot\)] (f1),
(ofn) -- [edge label = \(\cdot\)](fn)
};
\end{feynman}
\end{tikzpicture}
\end{equation}
\caption{Consecutive splitting $f_{(1\ldots m')} \to f_1 + \ldots + f_{m'}$ and $f\to (f_1 +\ldots +f_{m'}) +\ldots + f_m$ where the partons with momenta $\{\pN{},p_{m'+1},\ldots p_m\}$ become collinear to the direction $\pM{}$.}
\label{fig:nested_split2}
\end{figure}
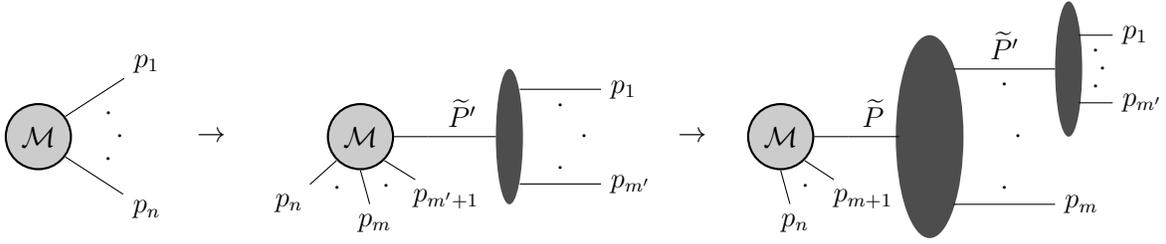

To be concrete, let us consider a collection of $m$ partons with flavour indices $\{f_1,\ldots, f_{m'},\ldots,f_m\}$ and momenta $\{p_1,\ldots,p_{m'},\ldots,p_m\}$, with $m'<m$. We always think of these partons as being part of an on-shell $n$-point amplitude $\mathcal M_{f_1 \ldots f_n}$ involving $(n-m)$ additional coloured partons. Our goal is to study the behaviour of the amplitude in the limit where $\{p_1,\ldots,p_{m'}\}$ become collinear to some lightlike direction $\pN{}$ and $\{\pN{},p_{m'+1},\ldots, p_m\}$ are collinear to another lightlike direction $\pM{}$. There are two different scenarios of how such a kinematic configuration can be reached, depending on the order in which the different collinear limits are taken. This is most conveniently understood by imagining the physical process where a parton with flavor $f$ and momentum $P = \sum_{i=1}^mp_i$ emits collinear radiation: 
\begin{enumerate}
        \item The parton with momentum $P$ and flavor $f$ splits into a set of collinear partons with momenta $\{p_{1},\ldots,p_m\}$ and flavors $\{f_{1},\ldots, f_m\}$, and the $m'$-parton subset becomes collinear to the direction $\pN{}$ (see fig.~\ref{fig:nested_split1}). We call this limit the \emph{iterated collinear limit}, and we denote it by $\mathscr{C}_{1\ldots m'}\mathscr{C}_{1 \ldots m}
    |\mathcal{M}_{f_1 \ldots f_n}(p_1,\ldots,p_n)|^2$. 
            \item  The partons with momenta $\{p_1,\ldots,p_{m'}\}$ and flavors $\{f_1,\ldots,f_{m'}\}$ become collinear to the direction $\pN{}$. Then, the parton with momentum $\pN{}$ and flavor $f_{(1\ldots m')}$, together with the partons with momenta $\{p_{m'+1},\ldots,p_m\}$ and flavors $\{f_{m'+1},\ldots, f_m\}$, become collinear to the direction $\pM{}$ (see fig.~\ref{fig:nested_split2}). We call this limit the \emph{strongly-ordered collinear limit}, and we denote it by $\mathscr{C}_{(1\ldots m')\ldots m}\mathscr{C}_{1\ldots m'}
    |\mathcal{M}_{f_1 \ldots f_n}(p_1,\ldots,p_n)|^2$.
\end{enumerate}
It is clear that these two limits describe the same region of phase space. Hence, the behaviours of the amplitude in the two limits must agree (because the value of the amplitude in a given point of phase cannot depend on how this point was approached), i.e., the strongly-ordered and iterated collinear limits of the amplitude must agree,
\begin{align}
    \mathscr{C}_{1\ldots m'}\mathscr{C}_{1 \ldots m}
    |\mathcal{M}_{f_1 \ldots f_n}(p_1,\ldots,p_n)|^2 =
    \mathscr{C}_{(1\ldots m')\ldots m}\mathscr{C}_{1\ldots m'}
    |\mathcal{M}_{f_1 \ldots f_n}(p_1,\ldots,p_n)|^2\,.
\label{eq:so_equivalence}
\end{align}
In the next section we will not distinguish them further, and only talk about the strongly-ordered limit. However, since the two limits are approached from different directions, i.e., they are computed from different kinematic parametrisations, we can exploit the fact that the limits agree in order to check whether our splitting amplitudes have the correct collinear sub-limits. This is discussed in more detail in App.~\ref{app:iterated_limit_of_split}.

\subsection{Parametrisation of the strongly-ordered collinear limit}
\label{sec:parametrisation_of_the_nested_collinear_limit}

In this section we give a precise definition of the strongly-ordered collinear limit.
We start by performing separate light-cone decompositions in each of the $m$- and $m'$-parton sets. For the $m$-parton set, we will use the notations and conventions of eq.~\eqref{eq:collinear_momenta}.
For the $m'$-parton subset we write
\begin{equation}
p^\mu_i = y_i \pN{\mu} + \Kb{i}^\mu - \frac{\Kb{i}^2}{2 y_i}\frac{n'^\mu}{n' \cdot \pN{}}\,, 
\quad i= 1,\ldots,m'\,,
\label{eq:Sudakov1.2}
\end{equation}
with $n'^2 = \pN{2} = \pN{}\cdot\Kb{i} = n'\cdot \Kb{i}=0$. 
The momenta $\pM{}$ and $\pN{}$ indicate the directions to which the partons in each set become collinear. We stress that at this point the lightcone directions $\pN{}$ and $n'$ in eq.~\eqref{eq:Sudakov1.2} are not related to the quantities $\pM{}$ and $n$ in eq.~\eqref{eq:collinear_momenta}. 
However, we may choose $n'=n$ without loss of generality. Indeed, if $\pM{} = E
(1,\vec{u})$ and $\pN{}=E'(1,\vec{u'})$ are given, we can choose any lightlike vectors $n$ and $n'$ such that $n\cdot \pM{}\neq0$ and $n'\cdot \pN{}\neq0$. For example, we may choose $n=(1,\vec v)$ and $n'=(1,\vec {v'})$, where $\vec{v}$ and $\vec{v'}$ are unit vectors. The choice of these unit vectors is arbitrary, as long as $\vec{u}\cdot\vec{v}\neq1$ and $\vec{u'}\cdot\vec{v'}\neq1$. It is then easy to see that we can always assume without loss of generality $\vec{v}=\vec{v'}$, i.e., $n=n'$. Let us also mention that, just like in the case of the ordinary collinear limit in Sec.~\ref{sec:setup}, we will always work in the axial gauge and we assume that the gauge vectors of all external and internal gluons is $n$.

With this setup, we can give a rigorous definition of the strongly-ordered collinear limit. Just like in the definition of the ordinary collinear limit in Sec.~\ref{sec:setup}, the vectors $\K{i}^\mu$ and $\Kb{i}^\mu$ parametrise the transverse distance to the planes spanned by $(\pM{},n)$ and $(\pN{},n)$, respectively. The strongly-ordered collinear limit where the $m'$-parton subset is more collinear then the $m$-parton set is then defined as the limit where both $\K{i}^\mu$ and $\Kb{i}^\mu$ approach zero, but the $\Kb{i}^\mu$ tend to zero faster than the $\K{i}^\mu$. We can implement the operation of taking this limit by a uniform rescaling of the transverse momenta in each collinear set by a different parameter,
\begin{align}
   &\K{i} \to \lambda \K{i}, \quad \Kb{i} \to \lambda' \Kb{i},
   \label{eq:so_rescaling}
\end{align}
and keeping the dominant singular terms of order $1/\lambda'^{\,2(m' -1)}\lambda^{2(m-m')}$ in the limit $\lambda, \lambda' \to 0$ with $ \lambda \gg \lambda'$. 

We conclude this section by discussing some properties of the parametrisation in eq.~\eqref{eq:Sudakov1.2}. First, the quantities $y_i$ and $\Kb{i}$ are not invariant under longitudinal boosts. Following the discussion in Sec.~\ref{sec:setup}, we can define
longitudinal-boost invariant quantities by (cf. eq.~(\ref{eq:boost})),
\begin{equation}
\zeta_l = \frac{y_l}{\sum_{j =1}^{m'} y_j} = \frac{p_l \cdot n}{P' \cdot n}\,,  \quad
\KTb{l}^\mu = \Kb{l}^\mu - \zeta_l \sum_{j=1}^{m'} \Kb{j}^\mu\,,
\quad l = 1,\ldots,m'\,,
\label{eq:boost_subset}
\end{equation}
where $P' = \sum_{i =1}^{m'} p_i$ is the parent momentum
of the $m'$-parton subset. 
Just like the variables defined in eq.~(\ref{eq:constraint}),
these new variables automatically satisfy the constraints,
\begin{equation}
\sum_{l=1}^{m'}\zeta_l = 1 \quad \textrm{and} \quad \sum_{l =1}^{m'} \KTb{l} =0.    
\end{equation}
From now on, we only work with these variables, and in order to avoid cluttering notation, we shall drop the tilde on the transverse momenta. Second, the momenta $p_i$, $i\le m'$, are in both sets, and so they admit lightcone decompositons according to both eq.~\eqref{eq:collinear_momenta} and eq.~\eqref{eq:Sudakov1.2}. It must therefore be possible to relate the variables $(\zeta_i,\Kb{i})$ and $(z_i,\K{i})$ for $i\le m'$. This is worked out in detail in App.~\ref{app:nested}. In particular, it is shown that $z_i$ and $\zeta_i$ are related through a simple ratio,
\begin{equation}
\zeta_i = \frac{z_i}{\sum_{j=1}^{m'} z_j}\,.
\end{equation}
We also show that the collinear directions are related by 
\begin{align}
\pN{\mu} = \pM{\mu}\sum_{i=1}^{m'} z_i  + \sum_{i=1}^{m'}\K{i}^\mu 
- \frac{\left(\sum_{i=1}^{m'}\K{i}\right)^2}{2\sum_{i=1}^{m'} z_i}\frac{n^\mu}{ n\cdot \pM{}}\,.
\label{eq:Sudakov_sublimit}
\end{align}
%

\subsection{Factorisation in the strongly-ordered collinear limits}
\label{sec:so_fac}
In this section we discuss how the tree-level squared amplitudes factorise in the strongly-ordered collinear limit. The corresponding factorisation formula follows immediately by iterating the collinear factorisation of the scattering amplitude in eq.~\eqref{eq:split_def},
{\begin{align} \label{eq:H_def}
\mathscr{C}_{(1\ldots m')\ldots m} & \mathscr{C}_{1\ldots m'} 
|\mathcal{M}_{f_1 \ldots f_n}(p_1,\ldots,p_n)|^2 = \left(\frac{2g_s^2\mu^{2\epsilon}}{s_{1\ldots m'}}\right)^{m'-1} \left(\frac{2g_s^2\mu^{2\epsilon}}{s_{[1\ldots m']\ldots m}}\right)^{m-m'} \nonumber\\
&\times 
\hat{P}^{hh'}_{f_1\ldots f_{m'}}\,\hat{H}^{hh';ss'}_{f_{(1\ldots m')}f_{m'+1}\ldots f_m}\,\mathcal{T}_{ff_{m+1}\ldots f_n}^{ss'}(\pM{},p_{m'+1},\ldots,p_n)\,,
\end{align}}
where 
\begin{align}
    s_{[1\ldots m']\ldots m} = (\pN{}+p_{m'+1}+\ldots +p_m)^2\,.
    \label{eq:s_square}
\end{align}
The functions $\hat{P}^{hh'}_{f_1\ldots f_{m'}}$ and $\mathcal{T}_{ff_{m+1}\ldots f_n}^{ss'}$ are the splitting amplitude and the helicity tensor introduced in Sec.~\ref{sec:setup}. The \emph{splitting tensor} $\hat{H}^{hh';ss'}_{f_{(1\ldots m')}f_{m'+1}\ldots f_m}$ is new. It is obtained by squaring the amplitude-level splitting amplitude without summing over the helicities of one of the partons in the collinear set (cf.~eq.~\eqref{eq:P_def}),
\beq\bsp
&\left(\frac{2g_s^2\mu^{2\epsilon}}{s_{[1\ldots m']\ldots m}}\right)^{m-m'}\,\hat{H}^{hh';ss'}_{f_{(1\ldots m')}f_{m'+1}\ldots f_m} \\
&\qquad = \frac1{\mathcal{C}_{f}}
\sum_{\substack{(s_{m'+1},\ldots,s_m)\\(c,c_{m'+1},\ldots,c_m)}}  {\bf Sp}_{f_{(1\ldots m')}f_{m'+1}\ldots f_m}^{c,c_{m'+1}\ldots c_m;s,h,s_{m'+1}\ldots s_m}\,\left[{\bf Sp}_{f_{(1\ldots m')}f_{m'+1}\ldots f_m}^{c,c_{m'+1}\ldots c_m;s',h',s_{m'+1}\ldots s_m}\right]^{\ast}\,,
\esp\eeq
where $\mathcal{C}_{f}$ is defined after eq.~(\ref{eq:P_def}).
Just like in Sec.~\ref{sec:setup} we suppress the dependence of all splitting amplitudes and tensors on their arguments. Due to the equivalence in eq.~\eqref{eq:so_equivalence}, the factorisation of the squared amplitude in the strongly-ordered limit can be cast in the form of a factorisation of the splitting amplitude itself,
\beq\label{eq:P_coll}
\mathscr{C}_{1\ldots m'} \hat{P}^{ss'}_{f_1\ldots f_{m}}= \left(\frac{s_{[1\ldots m']\ldots m}}{s_{1\ldots m'}}\right)^{m'-1} \,
\hat{P}^{hh'}_{f_1\ldots f_{m'}}\,\hat{H}^{hh';ss'}_{f_{(1\ldots m')}f_{m'+1}\ldots f_m}\,.
\eeq
By comparing eqs.~\eqref{eq:split_def} and~\eqref{eq:H_def}, it is easy to see that upon summing over the helicities $(h,h')$ the splitting tensor reduces to an ordinary splitting amplitude,
\beq\label{eq:H_to_P}
\delta^{hh'}\,\hat{H}^{hh';ss'}_{f_{(1\ldots m')}f_{m'+1}\ldots f_m} = \hat{P}^{ss'}_{f_{(1\ldots m')}f_{m'+1}\ldots f_m}\,.
\eeq

In this paper we only consider the situation where the parent parton with helicity indices $(s,s')$ is a quark. In that case additional simplifications occur. We start by discussing the case where the final-state parton with helicity indices $(h,h')$ is also a quark, $f_{(1\ldots m')} = q$. The splitting tensor may then involve three different tensor structures, 
\beq
\hat{H}^{hh';ss'}_{f_{(1\ldots m')}f_{m'+1}\ldots f_m} = {H}_1\,\delta^{hh'}\,\delta^{ss'} + {H}_2\,\delta^{hs}\,\delta^{h's'} + {H}_3\,\delta^{hs'}\,\delta^{h's}\,.
\eeq
Since $\hat{P}^{hh'}_{f_1\ldots f_{m'}}$ is proportional to $\delta^{hh'}$ in this case, only a specific linear combination of the coefficients $H_i$ enters eq.~\eqref{eq:P_coll}. Equation~\eqref{eq:H_to_P} then implies that in the case where the parent parton is a quark, we can cast eq.~\eqref{eq:P_coll} in a simpler form involving only unpolarised splitting amplitudes,
\beq\label{eq:P_coll_unpol}
\mathscr{C}_{1\ldots m'} \langle\hat{P}_{f_1\ldots f_{m}}\rangle= \left(\frac{s_{[1\ldots m']\ldots m}}{s_{1\ldots m'}}\right)^{m'-1} \,
\langle\hat{P}_{f_1\ldots f_{m'}}\rangle\,\langle\hat{P}_{qf_{m'+1}\ldots f_m}\rangle\,.
\eeq
If instead $f_{(1\ldots m')} = g$, helicity conservation on the parent quark line implies that
\beq
\label{eq:H_parentgluon}
\hat{H}^{hh';ss'}_{gf_{m'+1}\ldots f_m} = \delta^{ss'}\,
\hat{H}^{hh'}_{gf_{m'+1}\ldots f_m}\,.
\eeq
Just like for the ordinary splitting amplitude in Sec.~\ref{sec:setup}, it is convenient to work with Lorenz indices instead of helicity indices for the gluons.  
If we work in the axial gauge, the most general tensor structure is
\beq\bsp
&\hat{H}^{\alpha \beta}_{g f_{m'+1}\ldots f_m}  = 
d^{\alpha\beta}(\pM{},n)\,
A_{g f_{m'+1}\ldots f_m}
+\sum_{i,j =m'+1}^m \frac{\K{i}^\alpha\K{j}^\beta }{s_{[1\ldots m']\ldots m}} 
\,B_{ij,g f_{m'+1}\ldots f_m}.
\label{eq:helicity_tensor}
\esp\eeq
For the sake of clarity, we have omitted the dependence of the coefficients $A_{g f_{m'+1}\ldots f_m}$ and $B_{ij,g f_{m'+1}\ldots f_m}$ on their arguments. 
Unlike for the ordinary splitting amplitude, we cannot replace $d^{\alpha\beta}(\pM{},n)$ by $-g^{\alpha\beta}$, and so the splitting tensor depends on the gauge vector $n$, because the Lorentz indices of the splitting tensor are contracted with a splitting amplitude, which is not transverse (see the discussion at the 
end of Sec.~\ref{sec:setup}). The $n$-dependence cancels of course in every gauge-invariant physical quantity.

We have checked that our results for the quadruple splitting amplitudes have the correct behaviour in all strongly-ordered collinear limits, i.e., they satisfy eq.~\eqref{eq:P_coll_unpol} for $m'=2$ and $3$. The strongly-ordered limit of the quadruple splitting amplitudes involves the splitting tensors with two or three collinear particles in the final state. The relevant splitting tensors for two collinear partons are~\cite{Somogyi:2005xz}
\beq\bsp \label{eq:Hgq}
\hat{H}^{\alpha \beta}_{qg}\,& =\frac{1}{2}\,\delta^{hh'}\,\langle\hat{P}_{qg}\rangle\,,\\
\hat{H}^{\alpha \beta}_{g q} 
\,&=
C_F \left[
\frac{1}{2}\,(1-z)\,d^{\alpha \beta}(\pM{},n) - 2 \frac{z}{1-z}\frac{\K{}^\alpha \K{}^\beta}{\K{}^2}
\right]\,.
\esp\eeq
Similarly, the splitting tensors for three collinear partons are
\begin{align}
\label{eq:Hqff}
\hat{H}^{hh'}_{q f_1f_2}&= \frac{1}{2}\,\delta^{hh'}\,\langle\hat{P}_{qf_1f_2}\rangle\,,\\
\hat{H}^{\alpha \beta}_{g g q}&= 
C_F^2\, \hat{H}^{\alpha \beta\text{ (ab)}}_{g g q} 
+ C_F\, C_A\,  \hat{H}^{\alpha \beta \text{ (nab)}}_{g gq}\,.
\label{eq:Hggq}
\end{align}
The `abelian' and `non-abelian' pieces $\hat{H}^{\alpha \beta\text{ (ab)}}_{g g q}$ and $\hat{H}^{\alpha \beta \text{ (nab)}}_{g gq}$ 
have the tensor structure of eq.~(\ref{eq:helicity_tensor}), with coefficients
$A^{\text{ (ab)}}_{g g q}$, $B^{\text{ (ab)}}_{ij,g g q}$, $A^{\text{ (nab)}}_{g g q}$, $B^{\text{ (nab)}}_{ij,g g q}$ given in App.~\ref{app:3partonhelicity_tensor}.

\section{Tree-level soft currents}
\label{sec:iterated_soft-collinear}

Tree-level amplitudes do not only factorise in the limit where massless particles become collinear, but also when they are soft. We therefore expect that splitting amplitudes exhibit factorisation properties in the limit where some of the particles in the collinear set have vanishing energies. We start by discussing the soft limits of scattering amplitudes in general before turning to the soft limits of splitting amplitudes in subsequent sections.

We start by reviewing the behaviour of tree-level amplitudes as a massless parton becomes soft. We follow the notations and conventions of Sec.~\ref{sec:setup}. If a subset of $m$ partons become soft, i.e., have vanishing energy, the amplitude is divergent and the leading behaviour in the soft limit is described by the factorisation formula,
\beq\label{eq:soft_fac}
\mathscr{S}_{1\ldots m}\cM_{f_1\ldots f_n}^{c_1\ldots c_n;s_1\ldots s_n}(p_1,\ldots,p_n)= {\bf J}^{c_1\ldots c_m;s_1\ldots s_m} \cM_{f_{m+1}\ldots f_n}^{c_{m+1}\ldots c_n;s_{m+1}\ldots s_n}(p_{m+1},\ldots,p_n)\,,
\eeq
where $\mathscr{S}_{1\ldots m}$ denotes the operation of keeping only the leading divergent term in the limit. The {soft current} ${\bf J}^{c_1\ldots c_m;s_1\ldots s_m}$ is an operator on the colour space of the hard partons that maps the colour space of the hard particles to the total colour space of both the soft and the hard particles. It depends on the spins of the soft particles and the momenta of both the soft and hard particles. It also depends on the flavours of the soft particles, though we do not show this dependence explicitly.
The soft current has been computed at tree level for the emission of up to three soft partons~\cite{Bassetto:1984ik,Proceedings:1992fla,Catani:1999ss,Catani:2019nqv}. 

If we consider the current with the external polarisation vectors removed, {\it e.g.},
\beq\label{eq:gauge_invariance_soft}
{\bf J}^{c_1\ldots c_m;s_1\ldots s_m}\equiv \varepsilon^{s_1}_{\nu_1}(p_1,n)\dots\varepsilon^{s_m}_{\nu_m}(p_m,n)\, {\bf J}^{c_1\ldots c_m;\nu_1\ldots\nu_m}\,,
\eeq
it is expected that the soft current ${\bf J}^{c_1\ldots c_m;\nu_1\ldots\nu_m}$ be gauge invariant, in the sense that it vanishes when contracted with a soft gluon momentum, up to colour conservation in the hard amplitude,
\beq
p_k^{\nu_k}\,{\bf J}^{c_1\ldots c_k\ldots c_m}_{\nu_1\ldots \nu_k\ldots\nu_m} = 0\mod \sum_{i=m+1}^n{\bf T}_i^c=0\,.
\label{eq:gauge_invariance_soft_current}
\eeq
Equation~(\ref{eq:gauge_invariance_soft_current}) is known to hold for
$m=2$~\cite{Catani:1999ss} and for $m=3$~\cite{Catani:2019nqv}, and conjectured to hold for any $m$~\cite{Catani:2019nqv}.

The soft current for the emission of a single soft gluon reads,
\beq
{\bf J}^{c;s}(p_1) = \mu^{\eps}g_s\,\varepsilon_{\nu}^s(p_1,n)\,\sum_{i=2}^n{\bf T}_i^c\,\frac{p_i^{\nu}}{p_i\cdot p_1}\,,
\label{eq:singlesoftcurr}
\eeq
where the sum runs over the $n-1$ hard partons and ${\bf T}_i^c$ is the $\textrm{SU}(3)$ generator in the representation of parton $i$, i.e., $({\bf T}^c_i)_{ab} = {\bf T}^c_{a_ib_i} = T^c_{a_ib_i}$ if parton $i$ is a quark, $({\bf T}^c_i)_{ab} = - T^c_{b_ia_i}$ if it is an anti-quark, and $({\bf T}^c_i)_{ab} =i f^{a_icb_i}$ for a gluon.

The one-loop soft current is manifestly transverse, up to colour conservation in the hard amplitude,
\beq
p_1^{\nu}\,{\bf J}^c_{\nu} = \mu^{\eps}g_s\,\sum_{i=2}^n{\bf T}_i^c = 0\,.
\eeq
The transversality of the soft current implies that we can drop gauge-dependent terms in the polarisation sum of the gluons. For example, in the case of a single soft gluon, we have
\beq\bsp
\mathscr{S}_1&|\cM_{g_1f_2\ldots f_n}|^2 \\
&= \mu^{2\eps}g_s^2\sum_{j,k=2}^n\cS_{jk}(p_1)\left[\cM_{f_2\ldots f_n}^{c_2\ldots c'_j\ldots c_k\ldots c_{n};s_2\ldots s_{n}}\right]^{\ast}{\bf T}^{c_1}_{c'_jc_j} {\bf T}^{c_1}_{c_kc'_k}\cM_{f_2\ldots f_n}^{c_2\ldots c_j\ldots c'_k\ldots c_{n};s_2\ldots s_{n}}\,,
\label{eq:soft_amplitude_squared}
\esp\eeq
where we introduced the eikonal factor,
\beq
\cS_{jk}(p_1) = -\frac{2\,s_{jk}}{s_{1j}s_{1k}}\,.
\eeq
Note that colour matrices of different partons
commute, ${\bf T}_j\cdot {\bf T}_k = {\bf T}_k\cdot {\bf T}_j$, if $j\ne k$, and
\begin{equation}
    {\bf T}^{c_1}_{c'_jc_j}
    {\bf T}^{c_1}_{c_jc''_j}
    = ({\bf T}_j^2)_{c'_jc''_j}
    = C_j\, \delta_{c'_jc''_j}\,,
    \label{eq:colourtrivial}
\end{equation}
where ${\bf T}_j^2= C_j$ denotes the quadratic Casimir in the representation of parton $j$, i.e. $C_j=C_F$ if parton $j$ is a quark, $C_j=C_A$ if
it is a gluon.

Next, we examine the current for the emission of
a soft $q\bar{q}$ pair or of two soft gluons.
We consider first the case where a soft gluon splits into a quark of momentum $p_1$ and an antiquark of momentum $p_2$.
In the limit of the soft $q\bar{q}$ emission, the squared matrix element factorises as~\cite{Catani:1999ss}
\beq\bsp
&\mathscr{S}_{12}^{q\bar{q}}|\cM_{q_1\bar{q}_2f_3\ldots f_n}|^2 \\
&\,\, = (\mu^{2\eps}g_s^2)^2
\sum_{j,k=3}^n\cS_{jk}^{q\bar{q}}(p_1,p_2)
\left[\cM_{f_3\ldots f_n}^{c_3\ldots c'_j\ldots c_k\ldots c_{n};s_3\ldots s_{n}}\right]^{\ast}
{\bf T}^{c_{12}}_{c'_jc_j} 
{\bf T}^{c_{12}}_{c_kc'_k}\cM_{f_3\ldots f_n}^{c_3\ldots c_j\ldots c'_k\ldots c_{n};s_3\ldots s_{n}}\,,
\label{eq:qqdoublesoft_amplitude_squared}
\esp\eeq
where $c_{12}$ labels the colour of the gluon which splits
into the $q\bar{q}$ pair, and
\beq
\cS_{jk}^{q\bar{q}}(p_1,p_2) = \frac{4T_R}{s_{12}^2}
\frac{s_{1j}s_{2k}+s_{1k}s_{2j}-s_{jk}s_{12}}{ 
s_{j(12)}s_{k(12)} }\,,
\label{eq:qqdoublesoft_coeff}
\eeq
where 
\begin{equation}
    s_{i(12)}=s_{i1}+s_{i2}\,.
    \label{eq:s12}
\end{equation}

In  the case of two soft gluons of momenta $p_1$ and $p_2$, the factorisation of the squared matrix element can be separated into abelian and non-abelian parts~\cite{Catani:1999ss,Somogyi:2005xz},
\begin{equation}
    \mathscr{S}_{12}^{gg} |\cM_{g_1 g_2 f_3\ldots f_n}|^2=\left(\mathscr{S}_{12}^{gg,(ab)}+\mathscr{S}_{12}^{gg,(nab)}\right)|\cM_{g_1 g_2 f_3\ldots f_n}|^2.
\end{equation}
The abelian part is made out of a product of two single-gluon eikonal factors, with the colour correlations involving up to four hard partons,
\begin{align}
 \label{eq:doublesoftab}
    & \mathscr{S}_{12}^{gg,(ab)}|\cM_{g_1 g_2 f_3\ldots f_n}|^2  \\ 
    \nonumber
    & \,\,\, 
    = (\mu^{2\eps}g_s^2)^2\,\frac{1}{2}\sum_{i,j,k,l=3}^n
    \cS_{ik}(p_1)\cS_{jl}(p_2)
\left[\cM_{f_{3}\ldots f_n}^{c_3\ldots c'_i\ldots c'_j\ldots c_k\ldots c_\ell\ldots c_n;s_3\ldots s_{n}}\right]^{\ast}
\\ \nonumber & \,\,\, \times \left[ 
{\bf T}^{c_1}_{c'_ic_i} {\bf T}^{c_1}_{c_kc'_k} 
{\bf T}^{c_2}_{c'_jc_j} {\bf T}^{c_2}_{c_\ell  c'_\ell}
+ {\bf T}^{c_2}_{c'_jc_j} {\bf T}^{c_2}_{c_\ell  c'_\ell}
{\bf T}^{c_1}_{c'_ic_i} {\bf T}^{c_1}_{c_kc'_k} \right]
\cM_{f_{3}\ldots f_n}^{c_3\ldots c_i\ldots c_j\ldots
c'_k\ldots c'_\ell\ldots c_{n};s_3\ldots s_{n}}\,.
\end{align}
The non-abelian part, on the other hand, features the same colour correlations as the single-gluon and the 
soft-$q\bar{q}$-pair cases,
\beq\bsp
    &\mathscr{S}_{12}^{gg,(nab)}|\cM_{g_1 g_2 f_3\ldots f_n}|^2 
    = - (\mu^{2\eps}g_s^2)^2 C_A
    \\ &\,\,\,
    \sum_{j,k=3}^n\cS_{jk}^{gg}(p_1,p_2)|
    \left[\cM_{f_3\ldots f_n}^{c_3\ldots c'_j\ldots c_k\ldots c_{n};s_3\ldots s_{n}}\right]^{\ast}
    {\bf T}^{c_{12}}_{c'_jc_j}
    {\bf T}^{c_{12}}_{c_kc'_k}\cM_{f_3\ldots f_n}^{c_3\ldots c_j\ldots c'_k\ldots c_{n};s_3\ldots s_{n}}\,,
    \label{eq:doublesoftnab}
\esp\eeq
where 
\beq\bsp
    \cS_{jk}^{gg}(p_1,p_2)&\,=\cS_{jk}^{(s.o.)}(p_1,p_2)+4\frac{s_{j1}s_{k2}+s_{j2}s_{k1}}{s_{j(12)}s_{k(12)}}\left(\frac{1-\epsilon}{s_{12}^2}-\frac{1}{8}\cS_{jk}^{(s.o.)}(p_1,p_2)\right)\\
    &\,-\frac{8s_{jk}}{s_{12}s_{j(12)}s_{k(12)}}\,,
    \label{eq:doublesoftnabcoeff}
\esp\eeq
and
\begin{equation}
    \cS_{jk}^{(s.o.)}(p_1,p_2)=\cS_{jk}(p_2)\left(\cS_{j2}(p_1)+\cS_{k2}(p_1)-\cS_{jk}(p_1)\right)
\end{equation}
is the approximation of $\cS_{jk}(p_1,p_2)$ in the strongly-ordered limit.

\section{Single soft limit of tree-level splitting amplitudes}
\label{sec:singlesoftcollamp}

Having reviewed in Sec.~\ref{sec:iterated_soft-collinear} the soft limits of tree-level amplitudes in general, we now focus on the behaviour of splitting amplitudes in the limit where one gluon from the collinear set is soft. Since splitting amplitudes share many properties  of gauge-invariant amplitudes, we expect that the soft behaviour of splitting amplitudes can be encoded into universal factors closely related to the soft current reviewed in the previous section. 
In this section we derive a general formula that describes the soft limit of a splitting amplitude, and we show how we can recover the soft behaviour of triple collinear limits which was analysed in ref.~\cite{Somogyi:2005xz}.

Without loss of generality, we assume for now that the soft gluon is parton 1. The soft limit of the splitting amplitude $\hat{P}_{g_1f_2\ldots f_m}^{ss'}$ is defined as follows: we introduce a small parameter $\lambda$, and we perform the rescaling,
\beq
z_1\to \lambda z_1\,,\qquad  s_{1i}\to \lambda s_{1i}\,,\qquad k^\mu_{1\perp}\to \lambda k^\mu_{1\perp}\,,\qquad 1<i\le m\,.
\label{eq:softrescal}
\eeq
We expand the resulting function in $\lambda$ and only keep the leading pole in $\lambda$. As we now show, the coefficient of the leading pole is universal and described by a formula very reminiscent of the factorisation of the squared matrix element in eq.~(\ref{eq:soft_amplitude_squared}).

In order to obtain the factorisation of the splitting amplitude in the soft limit, we start from eq.~(\ref{eq:soft_amplitude_squared}) and take the collinear limit where the partons 1 through $m$ are collinear. In this limit, the amplitude on the right-hand side factorises according to eq.~\eqref{eq:split_def}. Next, we split the double sum in eq.~(\ref{eq:soft_amplitude_squared}) into four contributions:
\begin{enumerate}
    \item $m<j,k\le n$: Neither $j$ nor $k$ are in the collinear set. These terms do not contribute in the collinear limit, because the eikonal factor is not singular in the limit.
    \item $2\le j\le m<k\le n$: The eikonal factor has a simple pole in the collinear limit coming from $s_{1j}\to 0$. In the collinear limit, the eikonal factor reduces to
    \beq
    \mathscr{C}_{1\ldots m}\cS_{jk}(p_1) = -\frac{2\,z_j}{z_1\,s_{1j}}\,.
    \eeq
    \item $2\le k\le m<j\le n$: Similar to the previous case, with 
        \beq
    \mathscr{C}_{1\ldots m}\cS_{jk}(p_1) = -\frac{2\,z_k}{z_1\,s_{1k}}\,.
    \eeq
    \item $2\le j,k\le m$: Both invariants in the denominator of the eikonal factor are singular in the collinear limit. However, there is still only a simple pole, because the numerator also vanishes. Hence, these terms contribute at the same order as those with only one of the two particles in the collinear set.
\end{enumerate}
Putting everything together, we see that in the collinear limit eq.~(\ref{eq:soft_amplitude_squared}) becomes\footnote{For readability, we keep the spin dependence of all quantities implicit, except for the spin indices of the parent parton, in order to keep track of spin correlations.}
\begin{align}\label{eq:SC_derivation}
&\mathscr{C}_{1\ldots m}\mathscr{S}_1|\cM_{g_1f_2\ldots f_n}|^2= \\
\nonumber&\!= \mu^{2\eps}g_s^2\,\cT^{ss'}_{ff_{m+1}\ldots f_n} \frac1{\mathcal{C}_{f}}
\sum_{j,k=2}^m\cS_{jk}(p_1)\left[{\bf Sp}_{ff_{2}\ldots f_m}^{c,c_2\ldots c'_j\ldots c_k\ldots c_m;s'}\right]^{\ast} {\bf T}^{c_1}_{c'_jc_j}
{\bf T}^{c_1}_{c_kc_k'}{\bf Sp}_{ff_{2}\ldots f_m}^{c,c_2\ldots c_j\ldots c'_k\ldots c_{m};s}\\
\nonumber&\!-2\mu^{2\eps}g_s^2\sum_{k=2}^m\sum_{j=m+1}^n\!\frac{z_k}{z_1\,s_{1k}}\Big\{\!\!\left[\cM_{ff_{m+1}\ldots f_n}^{c,c_{m+1}\ldots c'_j\ldots c_{n};s'}\right]^{\ast}\left[{\bf Sp}_{ff_{2}\ldots f_m}^{c,c_2\ldots c_k\ldots c_m;s'}\right]^{\ast}
{\bf T}^{c_1}_{c'_jc_j}{\bf T}^{c_1}_{c_kc'_k}\\
\nonumber&\qquad\qquad \times{\bf Sp}_{ff_{2}\ldots f_m}^{c',c_2\ldots c'_k\ldots c_m;s}\cM_{ff_{m+1}\ldots f_n}^{c',c_{m+1}\ldots c_j\ldots c_{n};s}\\
\nonumber&\!\qquad +\left[\cM_{ff_{m+1}\ldots f_n}^{c,c_{m+1}\ldots c_j\ldots c_{n}}\right]^{\ast}\left[{\bf Sp}_{ff_{2}\ldots f_m}^{c,c_2\ldots c'_k\ldots c_m;s'}\right]^{\ast}
{\bf T}^{c_1}_{c'_kc_k}{\bf T}^{c_1}_{c_jc'_j}{\bf Sp}_{ff_{2}\ldots f_m}^{c',c_2\ldots c_k\ldots c_m;s}\cM_{ff_{m+1}\ldots f_n}^{c',c_{m+1}\ldots c'_j\ldots c_{n};s}\Big\}\,.
\end{align}
Next, we can use colour conservation for the hard amplitude in eq.~(\ref{eq:soft_amplitude_squared}), which asserts that
\beq\label{eq:colour_conservation}
\sum_{i=m+1}^n{\bf T}_i^{c_1} = -\sum_{i=2}^m{\bf T}_i^{c_1}\,.
\eeq
Applying this relation to the second and third lines in eq.~\eqref{eq:SC_derivation}, we find
\begin{align}\label{eq:SC_derivation_2}
&\mathscr{C}_{1\ldots m}\mathscr{S}_1|\cM_{g_1f_2\ldots f_n}|^2= \\
\nonumber&\!= \mu^{2\eps}g_s^2\,\cT^{ss'}_{ff_{m+1}\ldots f_n} \frac1{\mathcal{C}_{f}}
\sum_{j,k=2}^m\cS_{jk}(p_1)\left[{\bf Sp}_{ff_{2}\ldots f_m}^{c,c_2\ldots c'_j\ldots c_k\ldots c_m;s'}\right]^{\ast}{\bf T}^{c_1}_{c'_jc_j}
{\bf T}^{c_1}_{c_kc_k'}{\bf Sp}_{ff_{2}\ldots f_m}^{c,c_2\ldots c_j\ldots c'_k\ldots c_{m};s}\\
\nonumber&\!+2\mu^{2\eps}g_s^2\,\cT^{ss'}_{ff_{m+1}\ldots f_n} \frac1{\mathcal{C}_{f}}
\sum_{j,k=2}^m\frac{z_k}{z_1\,s_{1k}}\Big\{\!\!\left[{\bf Sp}_{ff_{2}\ldots f_m}^{c,c_2\ldots c'_j\ldots c_k\ldots c_m;s'}\right]^{\ast}{\bf T}^{c_1}_{c'_jc_j}
{\bf T}^{c_1}_{c_kc'_k}{\bf Sp}_{ff_{2}\ldots f_m}^{c,c_2\ldots c_j\ldots c'_k\ldots c_m;s} \\
\nonumber&\qquad\qquad\qquad\qquad\qquad\qquad
+\left[{\bf Sp}_{ff_{2}\ldots f_m}^{c,c_2\ldots c_j\ldots c'_k\ldots c_m;s'}\right]^{\ast}{\bf T}^{c_1}_{c'_kc_k}{\bf T}^{c_1}_{c_jc'_j}{\bf Sp}_{ff_{2}\ldots f_m}^{c,c_2\ldots c'_j\ldots c_k\ldots c_m;s}\Big\}\\
\nonumber&\!= \mu^{2\eps}g_s^2\,\cT^{ss'}
\frac1{\mathcal{C}_{f}}
\sum_{j,k=2}^mU_{jk;1}\,\left[{\bf Sp}_{ff_{2}\ldots f_m}^{c,c_2\ldots c'_j\ldots c_k\ldots c_m;s'}\right]^{\ast}{\bf T}^{c_1}_{c'_jc_j}
{\bf T}^{c_1}_{c_kc_k'}{\bf Sp}_{ff_{2}\ldots f_m}^{c,c_2\ldots c_j\ldots c'_k\ldots c_{m};s}\,,
\end{align}
where we defined
\beq
U_{jk;l}\equiv 2\left(  -\frac{s_{jk}}{s_{jl}s_{kl}} + \frac{z_k}{z_l s_{kl}} + \frac{z_j}{z_l s_{jl}} \right)\,.
\label{eq:ucoeff}
\eeq
We see that the hard matrix element completely factorises, and the soft gluon is only colour-correlated to the other collinear partons. This is a manifestation of colour-coherence: the cluster of collinear partons acts coherently as one single coloured object, and the hard partons cannot resolve its individual collinear constituents. Comparing eq.~\eqref{eq:SC_derivation_2} to eq.~\eqref{eq:factorisation}, we see that the splitting amplitude admits a factorisation very reminiscent of the squared matrix element in eq.~(\ref{eq:soft_amplitude_squared}),
\footnote{Note that in writing eq.~\eqref{eq:P_soft}, we have assumed that the soft limit and the collinear limit commute, which is certainly true in the strict limit.}
\beq\bsp\label{eq:P_soft}
\mathscr{S}_1&\left[
\left(\frac{2\mu^{2\eps}g_s^2}{s_{1\ldots m}}\right)^{m-1}
\hat{P}_{g_1f_2\ldots f_m}^{ss'}\right]\\
&\,= \mu^{2\eps}g_s^2 \frac1{\mathcal{C}_{f}}
\sum_{j,k=2}^mU_{jk;1}\,\left[{\bf Sp}_{ff_{2}\ldots f_m}^{c,c_2\ldots c'_j\ldots c_k\ldots c_m;s'}\right]^{\ast}{\bf T}^{c_1}_{c'_jc_j}
{\bf T}^{c_1}_{c_kc_k'}{\bf Sp}_{ff_{2}\ldots f_m}^{c,c_2\ldots c_j\ldots c'_k\ldots c_{m};s}\,.
\esp\eeq
Just like for the matrix element, the factorisation of soft emissions happens at the amplitude level and we have to keep track of colour correlations due to the soft emission. The main difference between eq.~(\ref{eq:soft_amplitude_squared}) and eq.~\eqref{eq:P_soft} is the fact that the eikonal factor is replaced by the quantity $U_{jk;l}$ defined in eq.~\eqref{eq:ucoeff}. 

Eq.~\eqref{eq:P_soft} generalises similar formulae found in the context of triple-collinear emissions in ref.~\cite{Somogyi:2005xz}, as we shall see in Sec.~\ref{sec:softqtoggq}. 
A few remarks are in order:
\begin{itemize}
    \item The factorisation in eq.~\eqref{eq:P_soft} is valid for an arbitrary number of collinear partons, and up to the colour correlations it is independent of the specific type or functional form of collinear amplitude.
    \item Up to replacing the quantity $U_{jk;l}$ with the appropriate factor, the factorisation in eq.~\eqref{eq:P_soft} is valid for any soft emission which is characterised by two-parton colour correlations, like e.g. a soft $q\bar{q}$ pair or the non-abelian part of the two-soft-gluon current, as we shall see in Sec.~\ref{sec:doublesoftlim}.
    \item For soft emissions characterised by colour correlations with four or more partons, eq.~\eqref{eq:P_soft} can be suitably generalised, as we shall see in Sec.~\ref{sec:doublesoftlim} in the case of the abelian part of the two-soft-gluon current.
\end{itemize}

When restricted to a specific collinear amplitude, the colour correlations of eq.~\eqref{eq:P_soft} can be further simplified. In the simplest case of a soft gluon
in a simple collinear limit, eq.~(\ref{eq:SC_derivation_2}) is reduced to
\begin{equation}
    \mathscr{C}_{12}\mathscr{S}_1|\cM_{g_1f_2\ldots f_n}|^2=
    \mu^{2\eps}g_s^2\,\cT^{ss'}_{ff_3\ldots f_n} 
    \frac1{\mathcal{C}_{f}} U_{22;1} \,
\left[{\bf Sp}_{ff_2}^{c,c'_2;s'}\right]^{\ast}
{\bf T}^{c_1}_{c'_2c_2}{\bf T}^{c_1}_{c_2c''_2}
{\bf Sp}_{ff_2}^{c,c''_2;s}\,.
\label{eq:s1c12}
\end{equation}
The colour algebra, eq.~(\ref{eq:colourtrivial}), is trivial. Further,
\begin{equation}
    \frac1{\mathcal{C}_{f}}
    \left[{\bf Sp}_{ff_2}^{c,c_2;s'}\right]^{\ast}
    {\bf Sp}_{ff_2}^{c,c_2;s} 
    = \delta_{ff_2} \delta^{ss'}\,.
    \label{eq:splitsquared}
\end{equation}
Then, using the explicit value of eq.~(\ref{eq:ucoeff})
for $j=k=2$, eq.~(\ref{eq:s1c12}) becomes
\begin{equation}\label{eq:soft_collinear_2}
    \mathscr{C}_{12}\mathscr{S}_1|\cM_{g_1f_2\ldots f_n}|^2=
    \mu^{2\eps}g_s^2\, \frac{4z_2}{z_1s_{12}}\, 
    C_2\, |\cM_{f_2\ldots f_n}|^2\,,
\end{equation}
which is the well-known soft-collinear limit. We can state eq.~\eqref{eq:soft_collinear_2} in the equivalent fashion,
\begin{equation}
    \mathscr{S}_1\left[\left(\frac{2\mu^{2\eps}g_s^2}{s_{12}}\right)\hat{P}_{g_1f_2}^{ss'}\right]=
    \mu^{2\eps}g_s^2\, \frac{4z_2}{z_1s_{12}}\, 
    C_2\,\delta^{ss'}\,.
\end{equation}

Next, we illustrate the factorisation in eq.~(\ref{eq:P_soft}) on the examples of triple and quadruple collinear splitting amplitudes.

\subsection{Single soft limit of a triple collinear limit}
\label{sec:softqtoggq}

The first non-trivial example of the factorisation in eq.~\eqref{eq:P_soft} is the single soft limit of a triple collinear limit. Let us consider a gluon, denoted as parton 1, becoming soft within the collinear limit of the process $f\to g_1 f_2 f_3$. In this case eq.~\eqref{eq:P_soft} becomes
\begin{align}\label{eq:P_s1_c123}
& \mathscr{S}_1\left[\left(\frac{2\mu^{2\eps}g_s^2}{s_{123}}\right)^{2}\hat{P}_{g_1f_2f_3}^{ss'}\right] 
=\mu^{2\eps}g_s^2 \frac1{\mathcal{C}_{f}} \\
\nonumber&\quad\times \left\{ U_{22;1} \,
\left[{\bf Sp}_{ff_2f_3}^{c,c'_2c_3;s'}\right]^{\ast}
{\bf T}^{c_1}_{c'_2c_2}{\bf T}^{c_1}_{c_2c''_2}
{\bf Sp}_{ff_2f_3}^{c,c''_2c_3;s} 
 + U_{23;1} 
\left[{\bf Sp}_{ff_2f_3}^{c,c'_2c_3;s'}\right]^{\ast}
{\bf T}^{c_1}_{c'_2c_2}{\bf T}^{c_1}_{c_3c'_3}
{\bf Sp}_{ff_2f_3}^{c,c_2c'_3;s} \right. \\
\nonumber&\quad \left.
+ U_{23;1} \left[{\bf Sp}_{ff_2f_3}^{c,c_2c'_3;s'}\right]^{\ast}
{\bf T}^{c_1}_{c'_3c_3} {\bf T}^{c_1}_{c_2c'_2}
{\bf Sp}_{ff_2f_3}^{c,c'_2c_3;s} 
 +U_{33;1} \left[{\bf Sp}_{ff_2f_3}^{c,c_2c'_3;s'}\right]^{\ast}
{\bf T}^{c_1}_{c'_3c_3} {\bf T}^{c_1}_{c_3c''_3}
{\bf Sp}_{ff_2f_3}^{c,c_2c''_3;s}
\right\}
\,,
\end{align}
where we used the symmetry of $U_{jk;1}$ under 
$j\leftrightarrow k$.
The functional form of eq.~(\ref{eq:P_s1_c123})
is characteristic of soft emissions which give rise to two-parton
colour correlations within a collinear structure made of two hard partons. 

In fact, using eq.~(\ref{eq:colourtrivial}) for $i=2,3$,
and the fact that colour charges of different partons
commute, ${\bf T}_2\cdot {\bf T}_3 = {\bf T}_3\cdot {\bf T}_2$, we can use colour conservation to write
\begin{equation}
    {\bf T}_2\cdot {\bf T}_3 = 
    \frac{{\bf T}_P^2-{\bf T}_2^2-{\bf T}_3^2}2\,,
    \label{eq:colourconserv}
\end{equation}
where ${\bf T}_P^2=C_P$ denotes the colour coefficient for emission by the parent $P$ of the collinear set.
We can therefore simplify eq.~(\ref{eq:P_s1_c123}),
\beq\bsp
\mathscr{S}_1&\left[\left(\frac{2\mu^{2\eps}g_s^2}{s_{123}}\right)^{2}\hat{P}_{g_1f_2f_3}^{ss'}\right] \\
&\, =\mu^{2\eps}g_s^2 \left[ U_{22;1} {\bf T}_2^2
+ U_{23;1} ({\bf T}_P^2-{\bf T}_2^2-{\bf T}_3^2) + U_{33;1} {\bf T}_3^2 \right]
\left(\frac{2\mu^{2\eps}g_s^2}{s_{23}}\right)\hat{P}_{f_2f_3}^{ss'}\,.
\label{eq:qqtofqq}
\esp\eeq
Using the specific values of $U_{jk;1}$, eq.~\eqref{eq:ucoeff}, and of the colour algebra,
for the process $q\to g_1 g_2 q_3$ we obtain
\beq\bsp
\mathscr{S}_1&\left[\left(\frac{2\mu^{2\eps}g_s^2}{s_{123}}\right)^{2}\hat{P}_{g_1g_2q_3}^{ss'}\right] \\
&\, = 2\mu^{2\eps}g_s^2\, \left[ \frac{2z_3}{z_1s_{13}} C_F
+ \left(\frac{s_{23}}{s_{12}s_{13}} + \frac{z_2}{z_1 s_{12}} -\frac{z_3}{z_1 s_{13}} \right) C_A \right] 
\left(\frac{2\mu^{2\eps}g_s^2}{s_{23}}\right)\hat{P}_{g_2q_3}^{ss'}\,.
\label{eq:g_1g_2q_3}
\esp\eeq
For the process $g\to g_1 \bar{q}_2 q_3,$ we obtain
\beq\bsp
\mathscr{S}_1&\left[\left(\frac{2\mu^{2\eps}g_s^2}{s_{123}}\right)^{2}\hat{P}_{g_1\bar{q}_2q_3}^{ss'}\right] \\
&\, = 2\mu^{2\eps}g_s^2\, \left[ 2 \frac{s_{23}}{s_{12}s_{13}} C_F
+ \left(- \frac{s_{23}}{s_{12}s_{13}} + \frac{z_2}{z_1 s_{12}} +\frac{z_3}{z_1 s_{13}} \right) C_A \right] 
\left(\frac{2\mu^{2\eps}g_s^2}{s_{23}}\right)\hat{P}_{\bar{q}_2q_3}^{ss'}\,.
\label{eq:g_1q_2q_3}
\esp\eeq
For the process $g\to g_1 g_2 g_3$, we obtain
\beq\bsp
\!\!\!\!\mathscr{S}_1\left[\left(\frac{2\mu^{2\eps}g_s^2}{s_{123}}\right)^{2}\!\hat{P}_{g_1g_2g_3}^{ss'}\right] 
 =2\mu^{2\eps}g_s^2\, \left( \frac{s_{23}}{s_{12}s_{13}} + \frac{z_2}{z_1 s_{12}} +\frac{z_3}{z_1 s_{13}} \right) C_A
\left(\frac{2\mu^{2\eps}g_s^2}{s_{23}}\right)\hat{P}_{g_2g_3}^{ss'}\,.
\label{eq:g_1g_2g_3}
\esp\eeq
Equations~\eqref{eq:g_1g_2q_3}, (\ref{eq:g_1q_2q_3})
and (\ref{eq:g_1g_2g_3}) are
in agreement with the single soft limit of a triple collinear limit computed in ref.~\cite{Somogyi:2005xz}.

\subsection{Single soft limit of the quadruple collinear limit}
We now consider the single soft limit of a quadruple collinear limit. We  denote the soft gluon as parton 1, within the quadruple collinear limit $f \to g_1 f_2 f_3 f_4$. The general single soft factorization formula (\ref{eq:P_soft}) becomes
\beq\bsp\label{eq:P_single_soft_quadruple}
\mathscr{S}_1&\left[\left(\frac{2\mu^{2\eps}g_s^2}{s_{1234}}\right)^{3}\hat{P}_{g_1f_2f_3 f_4}^{ss'}\right]\\
&\,= \mu^{2\eps}g_s^2 \frac1{\mathcal{C}_{f}}
\sum_{j,k=2}^4 
U_{jk;1}\,\left[{\bf Sp}_{ff_{2}f_{3}f_4}^{c,c_2\ldots c'_j\ldots c_k\ldots c_4;s'}\right]^{\ast}
{\bf T}^{c_1}_{c'_jc_j}{\bf T}^{c_1}_{c_kc_k'}{\bf Sp}_{ff_{2}f_{3}f_4}^{c,c_2\ldots c_j\ldots c'_k\ldots c_{4};s}\,.
\esp\eeq

The easiest non-trivial application of eq.~\eqref{eq:P_single_soft_quadruple} is the soft limit of the splitting $q\to g_1 \bar{q}'_2 q'_3 q_4$. 
Here the lower order process splitting amplitude is  ${\bf Sp}_{q\bar{q}^\prime_2 q^\prime_3  q_4}^{c,c_2c_3c_4;s}$ with the single colour structure $T^a_{c_3 c_2}T^a_{c_4 c}$. We can, therefore, explicitly compute the colour correlation structure of eq.~\eqref{eq:P_single_soft_quadruple} in terms of the single colour structure of the lower order splitting amplitude, i.e. make the replacements,
\beq
{\bf T}_2^2 = {\bf T}_3^2 = {\bf T}_4^2 = C_F\,,
\eeq
and
\begin{align}\label{eq:s1c1234_different_flavors_colour_correlations}
&{\bf T}_2\cdot {\bf T}_3 = \frac{C_A}{2}-C_F\,, \nonumber\\
&{\bf T}_2\cdot {\bf T}_4 = \frac{C_A}{2}-2C_F\,, \\
&{\bf T}_3\cdot {\bf T}_4 = 2C_F-C_A\,. \nonumber
\end{align}
Note that colour conservation implies
\beq
({\bf T}_2+{\bf T}_3+{\bf T}_4)^2 = {\bf T}_P^2 = C_F\,,
\eeq
so that one of the replacements in eq.~\eqref{eq:s1c1234_different_flavors_colour_correlations} can be derived from the other two. 
Inserting these expressions in eq.~\eqref{eq:P_single_soft_quadruple} we get 
\begin{align}\label{eq:singlesoft_final_qq_dif}
&\mathscr{S}_1\left[\left(\frac{2\mu^{2\eps}g_s}{s_{1234}}\right)^{3}\hat{P}_{g_1\bar{q}'_2q'_3 q_4}^{ss'}\right]=\mu^{2\eps}g_s^2 \bigg\{ U_{22;1} C_F + U_{33;1} C_F + U_{44;1}C_F 
\\
\nonumber&\, + 2 U_{23;1} \left(\frac{C_A}{2}-C_F\right) + 2 U_{24;1} \left(\frac{C_A}{2}-2C_F\right) + 2U_{34;1}(2C_F-C_A)  \bigg\}
\left(\frac{2\mu^{2\eps}g_s^2}{s_{234}}\right)^2\hat{P}_{\bar{q}'_2q'_3q_4}^{ss'} 
\\
\nonumber&\,
= \mu^{2\eps}g_s^2 \left[C_F B^{(q)}_{23,4}  +C_AA^{(q)}    \right]
\left(\frac{2\mu^{2\eps}g_s^2}{s_{234}}\right)^2\hat{P}_{\bar{q}'_2q'_3q_4}^{ss'}
\,,
\end{align}
where
\beq\bsp
\label{eq:single_soft_sublimit_quarks_Aq_Bq}
A^{(q)} &=  
\frac{4 z_2}{s_{12} z_1}
-\frac{2 z_3}{s_{13} z_1}
-\frac{2 z_4}{s_{14} z_1}
-\frac{2s_{23}}{s_{12} s_{13}}
-\frac{2 s_{24}}{s_{12} s_{14}}
+\frac{4 s_{34}}{s_{13}s_{14}}  \,,  \\
B_{ij,k}^{(qq)} &= 
\frac{4s_{ij}}{s_{1i} s_{1j}}
+\frac{8 s_{ik}}{s_{1i} s_{1k}}
-\frac{8 s_{jk}}{s_{1j} s_{1k}} 
-\frac{8 z_i}{s_{1i} z_1}
+\frac{8 z_j}{s_{1j} z_1}
+\frac{4 z_k}{s_{1k} z_1} \,.
\esp
\eeq
 
Next, we examine the soft limit of the splitting $q\to g\bar{q}qq$. The colour correlation is now more involved, since the lower order splitting ${\bf Sp}_{q\bar{q}_2 q_3  q_4}^{c,c_2c_3c_4;s}$ has two colour structures. We therefore decompose it into 
\beq
{\bf Sp}_{q\bar{q}_2 q_3  q_4}^{c, c_2c_3c_4;s} =  {\bf C}_1^{cc_2c_3c_4}{\bf F}^{(q\bar{q})s}_{1} + 
{\bf C}_2^{cc_2c_3c_4}
{\bf F}^{(q\bar{q})s}_{2} \,.
\eeq
with
\beq
{\bf C}_1^{cc_2c_3c_4} = T^a_{c_3c_2}T^a_{c_4c}\;\;\;\;,\;\;\;\; {\bf C}_2^{cc_2c_3c_4} = T^a_{c_4c_2}T^a_{c_3c}
\eeq
Upon squaring,
\beq\bsp
&
\sum_{{s_i}}\left[{\bf F}^{(q\bar{q})s'}_{1}\right]^\ast{\bf F}^{(q\bar{q})s}_{1} = \left(\frac{2\mu^{2\eps}g_s^2}{s_{234}}\right)^2 Q_{23,4} \delta^{ss'}\,,
\\
&
\sum_{{s_i}}\left[{\bf F}^{(q\bar{q})s'}_{2}\right]^\ast{\bf F}^{(q\bar{q})s}_{2} = \left(\frac{2\mu^{2\eps}g_s^2}{s_{234}}\right)^2 Q_{24,3}\delta^{ss'}\,,
\\
&
\sum_{{s_i}} \left[ \left({\bf F}^{(q\bar{q})s'}_{1}\right)^\ast 
{\bf F}^{(q\bar{q})s}_{2} + \text{c.c.}
\right]
= \left(\frac{2\mu^{2\eps}g_s^2}{s_{234}}\right)^2 \left(Q^{(id)}_{23,4}+Q^{(id)}_{24,3}\right)\delta^{ss'}\,,
\esp\eeq
where $Q_{ij,k}$ and $Q^{(id)}_{ij,k}$ are defined in eqs.~(\ref{eq:nnlo_splitting_kernels_1}) and (\ref{eq:nnlo_splitting_kernels_2}).
With these definitions, eq.~(\ref{eq:P_single_soft_quadruple}) becomes
\beq\bsp\label{eq:P_single_soft_quadruple_gqqq}
\mathscr{S}_1\left[\left(\frac{2\mu^{2\eps}g_s^2}{s_{1234}}\right)^{3}\hat{P}_{g_1f_2f_3 f_4}^{ss'}\right]
= \mu^{2\eps}g_s^2  \sum_{j,k=2}^4 
U_{jk;1}\,
\sum_{a,b=1}^2
\left[ {\bf F}^{(q\bar{q})s'}_{a} \right]^\ast {\bf F}^{(q\bar{q})s}_{b} \left[{\bf T}_j\cdot{\bf T}_k\right]_{a|b} \,,
\esp\eeq
where
\beq
\left[{\bf T}_j\cdot{\bf T}_k\right]_{a|b} = \frac1{\mathcal{C}_{f}} \left[{\bf C}_a^{cc_2 c_3 c_4}\right]_{c_j=c_{j'}}^\ast{\bf T}^{c_1}_{c_j' c_j} {\bf T}^{c_1}_{c_k c_k'}\left[{\bf C}_b^{cc_2 c_3 c_4}\right]_{c_k=c_{k'}} \,.
\eeq
As a concrete example of our notation,
\beq\bsp
\left[{\bf T}_2\cdot{\bf T}_3\right]_{1|2} &=  ({\bf C}_1^{cc'_2 c_3 c_4})^\ast{\bf T}^{c_1}_{c_2' c_2} {\bf T}^{c_1}_{c_3 c_3'} {\bf C}_2^{cc_2 c'_3 c_4} = 
T^a_{c'_2c_3}T^a_{cc_4} 
{\bf T}^{c_1}_{c_2' c_2} {\bf T}^{c_1}_{c_3 c_3'}
T^b_{c_4c_2}T^b_{c'_3c} 
\\
& = 
T^a_{c'_2c_3}T^a_{cc_4} 
{\bf T}^{c_1}_{c_2' c_2} {\bf T}^{c_1}_{c_3 c_3'}
T^b_{c_4c_2}T^b_{c'_3c} = -\Tr\left[ T^aT^bT^{c_1}T^aT^{c_1}T^b\right] \,.
\esp\eeq
It is, then  straightforward to compute the color correlations $\left[{\bf T}_j\cdot{\bf T}_k\right]_{a|b}$. We get
\beq\bsp\label{eq:s1c1234_equal_flavors_colour_correlations}
&\left[{\bf T}_j \cdot {\bf T}_j\right]_{11} = \left[{\bf T}_j \cdot {\bf T}_j\right]_{22} = C_F^2\,T_R\,  \,, \\
&\left[{\bf T}_2 \cdot {\bf T}_3\right]_{11} = \left[{\bf T}_2 \cdot {\bf T}_3\right]_{22} = -\left(C_F-\frac{C_A}{2}\right)\,C_F\, T_R\,\,, \\
&\left[{\bf T}_2 \cdot {\bf T}_4\right]_{11} = \left[{\bf T}_2 \cdot {\bf T}_4\right]_{22} = \left(\frac{C_A}{2}-2C_F\right)\, C_F\,T_R\, \,, \\
&\left[{\bf T}_3 \cdot {\bf T}_4\right]_{11} = \left[{\bf T}_3 \cdot {\bf T}_4\right]_{22} = (2C_F-C_A)\,C_F\,T_R\,  \,,\\
&\left[{\bf T}_j \cdot {\bf T}_j\right]_{12}  = C_F^2\left(C_F-\frac{C_A}{2}\right)\, \,,\\
&\left[{\bf T}_2\cdot {\bf T}_3\right]_{12} = -C_F \left(C_F-\frac{C_A}{2}\right)^2\, \,,\\
&\left[{\bf T}_2\cdot {\bf T}_4 \right]_{12} = -C_F \left(C_F-\frac{C_A}{2}\right)^2\, \,,\\
&\left[{\bf T}_3\cdot {\bf T}_4\right]_{12} = C_F\,(C_F-C_A)\,\left(C_F-\frac{C_A}{2}\right)\,\,.
\esp\eeq
We can then rewrite the soft limit in terms of $Q_{ij,k}$ and $Q^{(id)}_{ij,k}$ as 
\begin{align}\label{eq:singlesoft_final_qq_same}
&\mathscr{S}_1\left[
\Bigg(\frac{2\mu^{2\epsilon}\,{g_s}^2}{s_{1234}}\Bigg)^{3} \,
\hat{P}^{ss'}_{g_1 \bar{q}_2 q_3 q_4} \right]
= \delta^{ss'}g_s^2\mu^{2\epsilon}  \left(\frac{2g_s^2\mu^{2\epsilon}}{s_{234}}\right)^2 
\\
\nonumber& \qquad\qquad \times \left\{ \left[
C_F C_A T_R A^{(q)} +  C_F^2 T_R  B^{(q)}_{23,4}
\right] Q_{23,4} 
+\left[
C_FC_A T_R A^{(q)} +  C_F^2 T_R  B^{(q)}_{24,3}
\right]Q_{24,3} \right.
\\
\nonumber& \qquad\qquad\quad
\left.
+\left[ C_F\left(C_F-\frac{C_A}{2}\right)( C_A A^{(q)}+4C_F D^{(q)}) \right]\left(Q^{(id)}_{23,4}+Q^{(id)}_{24,3}\right) \right\}\,,
\end{align}
where $A^{(q)}$ and $B^{(q)}_{ij,k}$ are defined in eq.~(\ref{eq:single_soft_sublimit_quarks_Aq_Bq}), and 
\beq\bsp
&
D^{(q)}=-\frac{z_2}{s_{12} z_1}+\frac{z_3}{s_{13} z_1}+\frac{z_4}{s_{14} z_1}+\frac{s_{23}}{s_{12} s_{13}}+\frac{s_{24}}{s_{12} s_{14}}-\frac{s_{34}}{s_{13} s_{14}} \,.
   \esp
\eeq

Next, we consider the case
$q\to g_1 g_2 g_3 q_4$ with the gluon labeled 1 becoming soft. The underlying splitting amplitude can be written in terms of two colour structures, corresponding to the two colour orderings,
\beq\label{singlesoft_quadcol_gluons_splitting_amp_decomposition}
{\bf Sp}_{q g_2 g_3  q_4}^{c,c_2c_3c_4;s} = {\bf C}_{1}^{cc_2c_3c_4}{\bf F}_{1}^{(gg)} +
{\bf C}_{2}^{cc_2c_3c_4}{\bf F}_{2}^{(gg)}\,,
\eeq
with
\beq
{\bf C}_{1}^{cc_2c_3c_4} = (T^{c_2}T^{c_3})_{c_4c}\;\;\;\;,\;\;\;\;
{\bf C}_{2}^{cc_2c_3c_4} = (T^{c_3}T^{c_2})_{c_4c} \,.
\eeq
Upon squaring we have 
\beq\bsp\label{eq:singlesoft_quadcol_gluons_squares}
&
\sum_{{s_i}}\left[{\bf F}^{(gg)s'}_{1}\right]^\ast{\bf F}^{(gg)s}_{1} = \left(\frac{2\mu^{2\eps}g_s^2}{s_{234}}\right)^2 (Q^{(ab)}_{23,4}+2Q^{(nab)}_{23,4}+2Q^{(nab)}_{24,3}) \delta^{ss'}\,,
\\
&
\sum_{{s_i}}\left[{\bf F}^{(gg)s'}_{2}\right]^\ast{\bf F}^{(gg)s}_{2} = \left(\frac{2\mu^{2\eps}g_s^2}{s_{234}}\right)^2 (Q^{(ab)}_{24,3}+2Q^{(nab)}_{23,4}+2Q^{(nab)}_{24,3})\delta^{ss'}\,,
\\
&
\sum_{{s_i}} \left[ \left({\bf F}^{(gg)s'}_{1}\right)^\ast 
{\bf F}^{(gg)s}_{2} + \text{c.c.}
\right]
= -2 \left(\frac{2\mu^{2\eps}g_s^2}{s_{234}}\right)^2 \left(Q^{(nab)}_{23,4}+Q^{(nab)}_{24,3}\right)\delta^{ss'}\,,
\esp\eeq
where $Q^{(ab)}_{ij,k}$ and $Q^{(nab)}_{ij,k}$ are defined in eqs.~(\ref{eq:nnlo_splitting_kernels_3}) and (\ref{eq:nnlo_splitting_kernels_4}).
We can now write eq.~(\ref{eq:P_single_soft_quadruple}) as
\beq\bsp\label{eq:P_single_soft_quadruple_gggq}
\mathscr{S}_1\left[\left(\frac{2\mu^{2\eps}g_s^2}{s_{1234}}\right)^{3}\hat{P}_{g_1g_2g_3 q_4}^{ss'}\right]
= \mu^{2\eps}g_s^2  \sum_{j,k=2}^4 
U_{jk;1}\,
\sum_{a,b=1}^2
\left[ {\bf F}^{(gg)s'}_{a} \right]^\ast {\bf F}^{(gg)s}_{b} \left[{\bf T}_j\cdot{\bf T}_k\right]_{a|b}\,,
\esp\eeq
with
\beq\bsp
&\left[{\bf T}_2 \cdot {\bf T}_2\right]_{a | b} = C_A C_{a|b}\,,\\
&\left[{\bf T}_3 \cdot {\bf T}_3\right]_{a | b} = C_A C_{a|b}\,,\\
&\left[{\bf T}_4 \cdot {\bf T}_4\right]_{a | b} = C_F C_{a|b}\,,
\esp\eeq 
where 
\begin{equation}
C_{a|b} = \frac1{\mathcal{C}_{f}} \left[{\bf C}_a^{c c_2 c_3 c_4})\right]^\ast{\bf C}_b^{c c_2 c_3 c_4}\,,
\end{equation}
or explicitly,
\beq\bsp
C_{1|1} = C_{2|2} = C_F^2\,,\qquad
C_{1|2} = C_F\left(C_F-\frac{C_A}{2}\right) \,.
\esp\eeq
Moreover, 
\beq\bsp\label{eq:singlesoft_quadcol_colfactors}
& \left[{\bf T}_2 \cdot {\bf T}_3\right]_{1|1} = -\frac{C_A^2}{4}C_F\,,\\
& \left[{\bf T}_2 \cdot {\bf T}_3\right]_{2|2} = -\frac{C_A^2}{4}C_F\,,\\
& \left[{\bf T}_2 \cdot {\bf T}_3\right]_{1|2} = 0\,, \\
& \left[{\bf T}_2 \cdot {\bf T}_4\right]_{1|1} = \frac{1}{4}C_F\,,\\
& \left[{\bf T}_2 \cdot {\bf T}_4\right]_{2|2} = -\frac{C_A}{2}C_F^2\,,\\
& \left[{\bf T}_2 \cdot {\bf T}_4\right]_{1|2} = -\frac{C_A}{2}\left(C_F-\frac{C_A}{2}\right) \,,\\
& \left[{\bf T}_3 \cdot {\bf T}_4\right]_{1|1} = -\frac{C_A}{2}C_F^2\,,\\
& \left[{\bf T}_3 \cdot {\bf T}_4\right]_{2|2} = \frac{1}{4}C_F\,,\\
& \left[{\bf T}_3 \cdot {\bf T}_4\right]_{1|2} = -\frac{C_A}{2}\left(C_F-\frac{C_A}{2}\right)  \,.
\esp\eeq
Finally, replacing explicitly $U_{jk;1}$,
eq.~\eqref{eq:ucoeff},
and using eqs.~(\ref{eq:singlesoft_quadcol_gluons_squares}) and (\ref{eq:singlesoft_quadcol_colfactors}), 
we can write eq.~(\ref{eq:P_single_soft_quadruple_gggq}) as
\begin{align}\label{eq:singlesoft_final_gg}
&\mathscr{S}_1\left[\Bigg(\frac{2\mu^{2\epsilon}\,{g_s}^2}{s_{1234}}\Bigg)^{3}
\,\hat{P}^{ss'}_{g_1 g_2 g_3 q_4} \right]
\\
\nonumber&\,\, = \delta^{ss'} g_s^2\mu^{2\epsilon}  \left(\frac{2g_s^2\mu^{2\epsilon}}{s_{234}}\right)^2 
\left\{ 
\left[C_A C_F^2 A^{(gg)}_{23} +C_F C_A^2 B^{(gg)}+ C_F C^{(gg)}_{2} 
   +C_F^3D^{(gg)} \right]
Q^{(ab)}_{23,4}\right.
\\
\nonumber&\qquad\qquad\qquad\qquad\qquad\quad
+\left[ C_A C_F^2 A^{(gg)}_{32} 
   +C_F C_A^2 B^{(gg)}+C_F C^{(gg)}_{3} +C_F^3D^{(gg)} \right] Q^{(ab)}_{24,3}   \\
\nonumber& \qquad\qquad \left.  - \left[
C_A C_F^2 \left(2E^{(gg)}-D^{(gg)}\right)
   -C_A^2 C_F E^{(gg)}
   +2 C_F^3D^{(gg)}
\right]\left(Q^{(nab)}_{23,4}+Q^{(nab)}_{24,3}\right) \right\}
\,,
\end{align}
with
\beq\bsp
&A^{(gg)}_{ij}=\frac{4 z_i}{s_{1i} z_1}
+\frac{2 z_j}{s_{1j} z_1}
-\frac{2 z_4}{s_{14} z_1}
+\frac{2s_{j4}}{s_{1j} s_{14}} \,,\\
&B^{(gg)}=
-\frac{z_2}{s_{12} z_1}
-\frac{z_3}{s_{13}z_1}
+\frac{s_{23}}{s_{12} s_{13}} \,,\\
&C^{(gg)}_{i} = \frac{z_i}{s_{1i} z_1}+\frac{z_4}{s_{14}
   z_1}-\frac{s_{i4}}{s_{1i} s_{14}} \,,\\
&D^{(gg)}=\frac{4 z_4 }{s_{14} z_1} \,,\\
&E^{(gg)} = \frac{2 z_2}{s_{12} z_1}+\frac{2 z_3}{s_{13} z_1}-\frac{4 z_4}{s_{14} z_1}+\frac{2
   s_{24}}{s_{12} s_{14}}+\frac{2 s_{34}}{s_{13} s_{14}}\,.
\esp\eeq
Using eq.~(\ref{eq:softrescal}),
we have checked that a direct computation of the soft limit of the quadruple collinear splitting functions, that were presented in Sec.~\ref{sec:collinear}, agrees with the single soft factorization formulae for the quadruple collinear limits, eqs.~(\ref{eq:singlesoft_final_qq_dif}), (\ref{eq:singlesoft_final_qq_same}) and (\ref{eq:singlesoft_final_gg}). 

\section{Double soft limits of tree-level splitting amplitudes}
\label{sec:doublesoftlim}

Next, we consider the limit where a $q\bar{q}$ pair or two gluons from the collinear set are soft.
We take the soft $q\bar{q}$ pair or the two soft gluons
as partons 1 and 2. Like in eq.~(\ref{eq:softrescal}),
we introduce a small parameter $\lambda$, and perform the rescaling,
\beq\bsp
& z_1\to \lambda z_1\;\;,\;\; z_2\to \lambda z_2\;\;,\;\; k_{1\perp}^\mu \to \lambda k_{1\perp}^\mu
\;\;,\;\; k_{2\perp}^\mu \to \lambda k_{2\perp}^\mu \\
& s_{12}\to \lambda^2 s_{12}\;\;,\;\; s_{1i}\to \lambda s_{1i}\;\;\;,s_{2i} \to \lambda s_{2i}\,,\qquad 2<i\le m\,.
\label{eq:doublesoftrescal}
\esp\eeq
We expand the resulting function in $\lambda$ and only keep the leading pole in $\lambda$, which is $\mathcal{O}(\lambda^{-4})$.

\subsection{Soft $q\bar{q}$ pair}

For the soft $q\bar{q}$ pair, whose current in eq.~\eqref{eq:qqdoublesoft_amplitude_squared} features two-parton
colour correlations like the single gluon soft current, the derivation of the soft limit of a tree-level splitting amplitude is the similar to the one of Sec.~\ref{sec:singlesoftcollamp}.
Namely, we split the double sum of eq.~(\ref{eq:qqdoublesoft_amplitude_squared}) into four contributions,
and using colour conservation for the hard amplitude in eq.~\eqref{eq:soft_amplitude_squared},
\beq\label{eq:colour_conservation2}
\sum_{i=m+1}^n{\bf T}_i^{c_{12}} = -\sum_{i=3}^m{\bf T}_i^{c_{12}}\,,
\eeq
we arrive at an expression which is formally similar to
eq.~(\ref{eq:P_soft}),
\beq\bsp\label{eq:P_qqdoublesoft}
\mathscr{S}_{12}^{q\bar{q}}
&\left[\left(\frac{2\mu^{2\eps}g_s^2}{s_{1\ldots m}}\right)^{m-1}
\hat{P}_{q_1\bar{q}_2f_3\ldots f_m}^{ss'}\right]\\
&\,= (\mu^{2\eps}g_s^2)^2 \frac1{\mathcal{C}_{f}}
\sum_{j,k=3}^m U_{jk;12}^{q\bar{q}}\,\left[{\bf Sp}_{ff_{3}\ldots f_m}^{c,c_3\ldots c'_j\ldots c_k\ldots c_m;s'}\right]^{\ast}{\bf T}^{c_{12}}_{c'_jc_j}
{\bf T}^{c_{12}}_{c_kc_k'}{\bf Sp}_{ff_{3}\ldots f_m}^{c,c_3\ldots c_j\ldots c'_k\ldots c_{m};s}\,,
\esp\eeq
where
\beq\bsp
U_{jk;12}^{q\bar{q}}(p_1,p_2) &= \frac{4T_R}{s_{12}^2} \biggl(
\frac{s_{1j}s_{2k}+s_{1k}s_{2j}-s_{jk}s_{12}}{ 
s_{j(12)}s_{k(12)} } 
- \frac{z_1s_{2j}+z_2s_{1j}-z_js_{12}}{(z_1+z_2)s_{j(12)}} \\
& \qquad\qquad
- \frac{z_1s_{2k}+z_2s_{1k}-z_ks_{12}}{(z_1+z_2)s_{k(12)}}
+ \frac{2z_1z_2}{(z_1+z_2)^2}
\biggr)\,,
\label{eq:qqdoublesoft_coeffu}
\esp\eeq
where $s_{j(12)},\,s_{k(12)}$ are defined in eq.~(\ref{eq:s12}).
The first term on the right-hand side of
eq.~(\ref{eq:qqdoublesoft_coeffu}) stems from 
$j$ and $k$ being in the collinear set,
$3\le j,k\le m$. Then all the invariants in the eikonal factor $\cS_{jk}^{q\bar{q}}(p_1,p_2)$ in eq.~\eqref{eq:qqdoublesoft_coeff} are singular in the collinear limit, and $\cS_{jk}^{q\bar{q}}(p_1,p_2)$ contributes as such.
The second and third terms of
eq.~\eqref{eq:qqdoublesoft_coeffu} stem from either
$j$ being in the collinear set and $k$ outside,
$3\le j\le m<k\le n$, or viceversa, $3\le k\le m<j\le n$. They are obtained from the eikonal factor $\cS_{jk}^{q\bar{q}}(p_1,p_2)$ in eq.~\eqref{eq:qqdoublesoft_coeff} using the rescaling in eq.~\eqref{eq:doublesoftrescal}.
The last term in eq.~(\ref{eq:qqdoublesoft_coeffu}) stems from neither $j$ nor $k$ being in the collinear set, $m<j,k\le n$. However, at variance with the single
soft limit analysed in Sec.~\ref{sec:singlesoftcollamp},
the first two terms of the eikonal factor in eq.~(\ref{eq:qqdoublesoft_coeff}) contribute in the collinear limit, when $m<j,k\le n$.

We display how the factorisation in eq.~\eqref{eq:P_qqdoublesoft} works in the simplest case of a soft $q\bar{q}$ pair in the triple collinear splitting, $q\to q'_1 \bar{q}'_2 q_3$ or $g\to q_1 \bar{q}_2 g_3$.
The double sum in eq.~(\ref{eq:P_qqdoublesoft}) is reduced to
a single factor, $j=k=3$, with ${\bf T}_3^2=C_3$,
where $C_3=C_F$ if particle 3 is a quark or $C_3=C_A$ if particle 3 is a gluon.
Then, using eq.~(\ref{eq:splitsquared}) and
the explicit value of eq.~(\ref{eq:qqdoublesoft_coeffu})
for $j=k=3$,  we can write
\begin{equation}
  \!\! \!\!  \mathscr{S}_{12}^{q\bar{q}}\left[\left(\frac{2\mu^{2\eps}g_s^2}{s_{123}}\right)^{2}\hat{P}_{q_1\bar{q}_2f_3}^{ss'}\right] =
    (\mu^{2\eps}g_s^2)^2\,
    \frac{8T_R}{s_{12}^2}
    \left( \frac{z_3}{z_1+z_2} \frac{s_{12}}{s_{3(12)}}
    - \frac{(z_1s_{23}-z_2s_{13})^2}{(z_1+z_2)^2 s_{3(12)}^2}
    \right)\, C_3\, \delta^{ss'}\,,
\end{equation}
or equivalently,
\beq\bsp
    \mathscr{C}_{123} &\mathscr{S}_{12}^{q\bar{q}} |\cM_{q_1\bar{q}_2f_3\ldots f_n}|^2 =\\
  &  (\mu^{2\eps}g_s^2)^2\,
    \frac{8T_R}{s_{12}^2}
    \left( \frac{z_3}{z_1+z_2} \frac{s_{12}}{s_{3(12)}}
    - \frac{(z_1s_{23}-z_2s_{13})^2}{(z_1+z_2)^2 s_{3(12)}^2}
    \right)\, C_3\,
    |\cM_{f_3\ldots f_n}|^2\,,
    \label{eq:c123s2qq}
\esp\eeq
in agreement with ref.~\cite{Somogyi:2005xz}. 

\subsection{Soft $q\bar{q}$ pair limit of a quadruple collinear limit}
\label{sec:softqqtoqqff}

The first non-trivial example of the factorisation in eq.~\eqref{eq:P_qqdoublesoft} is the soft $q\bar{q}$ pair limit of a quadruple collinear limit. We denote the soft $q\bar{q}$ pair as partons 1 and 2 within the collinear limit of the process $f\to q_1 \bar{q}_2 f_3 f_4$.
We stated in Sec.~\ref{sec:softqtoggq} that
the functional form of eq.~(\ref{eq:P_s1_c123})
is characteristic of soft emissions which give rise to two-parton
colour correlations within a collinear structure made of two hard partons. 
Thus, it applies to soft $q\bar{q}$ emission
within a quadruple collinear amplitude as well, up to
replacing  in eq.~(\ref{eq:qqtofqq}) the factors $U_{jk;1}$ from eq.~(\ref{eq:ucoeff}) with $U_{jk;12}^{q\bar{q}}$ from eq.~(\ref{eq:qqdoublesoft_coeffu}),
\beq\bsp
\mathscr{S}_{12}^{q\bar{q}}&\left[\left(\frac{2\mu^{2\eps}g_s^2}{s_{1234}}\right)^{3}\hat{P}_{q_1\bar{q}_2f_3f_4}^{ss'}\right] \\
&\, = (\mu^{2\eps}g_s^2)^2\, 
\left[ U_{33;12}^{q\bar{q}} {\bf T}_3^2
+ U_{34;12}^{q\bar{q}} ({\bf T}_P^2-{\bf T}_3^2-{\bf T}_4^2) + U_{44;12}^{q\bar{q}} {\bf T}_4^2 \right]
\left(\frac{2\mu^{2\eps}g_s^2}{s_{34}}\right)\hat{P}_{f_3f_4}^{ss'}\,,
\label{eq:qqtoffqq}
\esp\eeq
where ${\bf T}_P^2=C_P$ denotes the colour coefficient of the parent $P$ of the collinear set, and $U_{34;12}^{q\bar{q}}$ is obtained readily from
eq.~(\ref{eq:qqdoublesoft_coeffu}). We mention  $U_{33;12}^{q\bar{q}}$ and $U_{44;12}^{q\bar{q}}$ 
explicitly since they have a rather compact form, which
has been used in eq.~(\ref{eq:c123s2qq}),
\begin{equation}
    U_{ii;12}^{q\bar{q}} =
    \frac{8T_R}{s_{12}^2}
    \left( \frac{z_i}{z_1+z_2} \frac{s_{12}}{s_{i(12)}}
    - \frac{(z_1s_{2i}-z_2s_{1i})^2}{(z_1+z_2)^2 s_{i(12)}^2}
    \right)\,, \qquad i=3, 4\,.
\end{equation}
As in Sec.~\ref{sec:softqtoggq}, through eq.~(\ref{eq:qqtoffqq}) we may describe
three processes, which we list with their corresponding
colour algebra,
\begin{eqnarray}
q\to q'_1 \bar{q}'_2 g_3 q_4\,, 
& \quad q\to q_1 \bar{q}_2 g_3 q_4\,,
&\qquad {\bf T}_3^2=C_A\,,\quad {\bf T}_4^2={\bf T}_P^2=C_F\,, \nonumber\\
g\to q_1 \bar{q}_2 q'_3 \bar{q}'_4\,, 
& \quad g\to q_1 \bar{q}_2 q_3 \bar{q}_4\,,
&\qquad
{\bf T}_3^2={\bf T}_4^2=C_F\,,\quad {\bf T}_P^2=C_A\,, \\
g\to q_1 \bar{q}_2 g_3 g_4\,, & &\qquad
{\bf T}_3^2={\bf T}_4^2={\bf T}_P^2=C_A\,. \nonumber
\label{eq:twohardalg}
\end{eqnarray}
We have checked that, if we use the color factors in the first line of eq.~\eqref{eq:twohardalg}, then our results for the splitting amplitudes $\hat{P}_{q\bar{q}gq}^{ss'}$ and $\hat{P}_{q'\bar{q}'gq}^{ss'}$ are consistent with the soft factorisation in eq.~\eqref{eq:qqtoffqq}.

\subsection{Two soft gluons}

The derivation of the two-soft gluon limit of the 
tree-level splitting amplitudes is more involved, since
the two-soft gluon current features also four-parton
colour correlations in the abelian part in eq.~\eqref{eq:doublesoftab}.

We start from the non-abelian part in eq.~(\ref{eq:doublesoftnab}), which features only two-parton colour correlations. 
We split the double sum of eq.~(\ref{eq:doublesoftnab}) into four contributions, and repeat the analysis done
for the soft $q\bar{q}$ pair. Like in that case,
when neither $j$ nor $k$ are in the collinear set, there is a contribution from the double pole in $s_{12}$ of eq.~(\ref{eq:doublesoftnabcoeff}).
We obtain then an expression which is yet again similar to eq.~(\ref{eq:P_soft}),
\begin{align}\label{eq:P_ggnabdoublesoft}
&\mathscr{S}_{12}^{gg (nab)}
\left[\left(\frac{2\mu^{2\eps}g_s^2}{s_{1\ldots m}}\right)^{m-1}
\hat{P}_{g_2g_2f_3\ldots f_m}^{ss'}\right]\\
\nonumber&\quad = - (\mu^{2\eps}g_s^2)^2 \frac1{\mathcal{C}_{f}}\, C_A 
\sum_{j,k=3}^m U_{jk;12}^{gg (nab)}\,\left[{\bf Sp}_{ff_{3}\ldots f_m}^{c,c_3\ldots c'_j\ldots c_k\ldots c_m;s'}\right]^{\ast}{\bf T}^{c_{12}}_{c'_jc_j}
{\bf T}^{c_{12}}_{c_kc_k'}{\bf Sp}_{ff_{3}\ldots f_m}^{c,c_3\ldots c_j\ldots c'_k\ldots c_{m};s}\,,
\end{align}
where
\begin{equation}
    U_{jk;12}^{gg (nab)} = 
    \cS_{jk}^{gg}(p_1,p_2) - \cS_{j}^{gg}(p_1,p_2) - \cS_{k}^{gg}(p_1,p_2)
    + 4\, \frac{2z_1z_2}{(z_1+z_2)^2} \frac{1-\epsilon}{s_{12}^2}\,,
    \label{eq:U_ggnabdoublesoft}
\end{equation}
with

\begin{align}
    \label{eq:csieq}
&\cS_{i}^{gg}(p_1,p_2) =\\ 
\nonumber&\,\, = \cS_{i}^{(s.o.)}(p_1,p_2)
+4\frac{z_2s_{i1}+z_1s_{i2}}{(z_1+z_2)s_{i(12)}}\left(\frac{1-\epsilon}{s_{12}^2}-\frac{1}{8}\cS_{i}^{(s.o.)}(p_1,p_2)\right)
    -\frac{4}{s_{12}}\frac{2z_i}{(z_1+z_2)s_{i(12)}}\,,
\end{align}
and
\begin{equation}
\cS_{i}^{(s.o.)}(p_1,p_2) =
- \frac{2 z_i}{z_2 s_{i2}} \left( -\frac{2s_{i2}}{s_{i1}s_{12}}
- \frac{2 z_2}{z_1 s_{12}} + \frac{2 z_i}{z_1 s_{i1}} \right)\,,
\end{equation}

As regards the abelian part of the two-soft gluon current,
the quadruple sum in eq.~(\ref{eq:doublesoftab}) may be split into
several contributions:
\begin{enumerate}
    \item $m<i,j,k,\ell\le n$. None of $i$, $j$, $k$, $\ell$ are in the collinear set. These terms do not contribute in the collinear limit, because the eikonal factors are not singular in the limit.
    \item $3\le i\le m< j,k,\ell\le n$, and likewise the other three
    cases where only one index runs in the collinear set.
    These terms do not contribute in the collinear limit, because one eikonal factor is not singular in the limit.
    \item $3\le i,k\le m< j,\ell\le n$, and likewise the other
    case where only the pair $(j,\ell)$ runs through the collinear set.
    These terms do not contribute in the collinear limit, because one eikonal factor is not singular in the limit.
    \item $3\le i,j\le m < k,\ell\le n$, and likewise the other three
    cases where one index from the pair $(i,k)$ and one index from
    the pair $(j,\ell)$ run through the collinear set.
    Each eikonal factor has a simple pole in the collinear limit.
    For example, for $3\le i,j\le m < k,\ell\le n$,
    the eikonal factors reduce to
    \beq
    \mathscr{C}_{1\ldots m}\cS_{ik}(p_1)\cS_{j\ell}(p_2) = \frac{2\,z_i}{z_1\,s_{1i}}\, \frac{2\,z_j}{z_2\,s_{2j}}\,,
    \eeq
    and likewise for the other three cases.
    \item $3\le i,j,k\le m < \ell\le n$, and likewise the other three
    cases where one index is outside of the collinear set. For $3\le i,j,k\le m < \ell\le n$, we obtain
    \beq
    \mathscr{C}_{1\ldots m}\cS_{ik}(p_1)\cS_{j\ell}(p_2) = - \frac{2\,z_j}{z_2\,s_{2j}}\, \cS_{ik}(p_1),
    \eeq
    and likewise for the other three cases.
    \item $3\le i,j,k, \ell\le m$. Each eikonal factor behaves like in the analogous case discussed in Sec.~\ref{sec:singlesoftcollamp} for the
    single soft limit.
\end{enumerate}
Putting together the contributions outlined above,
and using colour conservation for the hard amplitude in eq.~(\ref{eq:soft_amplitude_squared}) as in eq.~(\ref{eq:colour_conservation}),
the coefficient of the abelian part in eq.~(\ref{eq:doublesoftab}) of the double soft gluon limit becomes, as expected, the product of two
single gluon coefficients $U_{jk;l}$, defined in eq.~(\ref{eq:ucoeff}),
\begin{align}\label{eq:P_ggdoublesoft_ab}
& \mathscr{S}_{12}^{gg, (ab)}
\left[\left(\frac{2\mu^{2\eps}g_s^2}{s_{1\ldots m}}\right)^{m-1}\hat{P}_{g_1g_2f_3\ldots f_m}^{ss'}\right]
\\ \nonumber &\,\,\, 
= (\mu^{2\eps}g_s^2)^2 \frac1{\mathcal{C}_{f}}\, 
\frac1{2} \sum_{i,j,k,\ell=3}^m U_{ik;1}\, U_{j\ell;2}
\left[{\bf Sp}_{ff_{3}\ldots f_m}^{c,c_3\ldots c'_i\ldots c'_j\ldots c_k\ldots c_\ell\ldots c_m;s'}\right]^{\ast}
\\ \nonumber & \,\,\, \times \left[ 
{\bf T}^{c_1}_{c'_ic_i} {\bf T}^{c_1}_{c_kc'_k} 
{\bf T}^{c_2}_{c'_jc_j} {\bf T}^{c_2}_{c_\ell  c'_\ell}
+ {\bf T}^{c_2}_{c'_jc_j} {\bf T}^{c_2}_{c_\ell  c'_\ell}
{\bf T}^{c_1}_{c'_ic_i} {\bf T}^{c_1}_{c_kc'_k} \right]
{\bf Sp}_{ff_{3}\ldots f_m}^{c,c_3\ldots c_i\ldots c_j\ldots
c'_k\ldots c'_\ell\ldots c_{m};s}\,.
\end{align}
The sum of eqs.~(\ref{eq:P_ggnabdoublesoft}) and 
(\ref{eq:P_ggdoublesoft_ab}), 
\beq\bsp
    \mathscr{S}_{12}^{gg}&
\left[\left(\frac{2\mu^{2\eps}g_s^2}{s_{1\ldots m}}\right)^{m-1}\hat{P}_{g_1g_2f_3\ldots f_m}^{ss'}\right]=\\
  &  = \left( \mathscr{S}_{12}^{gg, (ab)} + \mathscr{S}_{12}^{gg, (nab)}\right)
\left[\left(\frac{2\mu^{2\eps}g_s^2}{s_{1\ldots m}}\right)^{m-1}\hat{P}_{g_1g_2f_3\ldots f_m}^{ss'}\right]\,,
\esp\eeq
yields the double-soft gluon limit of an $m$-parton collinear amplitude.

In the simplest case of two soft gluons in a triple collinear splitting, $q\to g_1 g_2 q_3$ or $g\to g_1 g_2 g_3$, the quadruple sum in eq.~(\ref{eq:P_ggdoublesoft_ab}) is reduced to a single factor, $i=j=k=\ell=3$. Thus, we can write
\begin{equation}\bsp
    & \mathscr{C}_{123} \mathscr{S}_{12}^{gg, (ab)} |\cM_{g_1g_2f_3\ldots f_n}|^2 =
    (\mu^{2\eps}g_s^2)^2\,
    \cT^{ss'}_{ff_4\ldots f_n}\, \frac1{\mathcal{C}_{f}}\,
    \frac1{2}\, U_{33;1}\, U_{33;2}\\
&\,\,\, \times 
\left[ {\bf Sp}_{ff_3}^{c,c'_3;s'}\right]^{\ast}
\left[ 
\left({\bf T}^{c_1}{\bf T}^{c_1}{\bf T}^{c_2}{\bf T}^{c_2}\right)_{c'_3 c_3}
+ \left({\bf T}^{c_2}{\bf T}^{c_2}{\bf T}^{c_1}{\bf T}^{c_1}\right)_{c'_3 c_3}
\right] 
{\bf Sp}_{ff_3}^{c,c_3;s}\,.
\esp\end{equation}
Using the colour algebra in eq.~(\ref{eq:colourtrivial}),
the products of four ${\bf T}$'s reduce to
\begin{equation}
    \left({\bf T}^{c_1}{\bf T}^{c_1}{\bf T}^{c_2}{\bf T}^{c_2}\right)_{c'_3 c_3}
+ \left({\bf T}^{c_2}{\bf T}^{c_2}{\bf T}^{c_1}{\bf T}^{c_1}\right)_{c'_3 c_3} = 
    2 ({\bf T}^2_3)^2_{c'_3c_3}
    = 2 C_3^2 \delta_{c'_3c_3} \,,
\end{equation}
where $C_3=C_F$ if particle 3 is a quark or $C_3=C_A$ if it is a gluon.
Using then eq.~(\ref{eq:splitsquared}) and
the explicit value of eq.~(\ref{eq:ucoeff})
for $i=k=3$ and $j=\ell=3$, we can write
\begin{equation}
    \mathscr{C}_{123} \mathscr{S}_{12}^{gg, (ab)} |\cM_{g_1g_2f_3\ldots f_n}|^2 =
    (\mu^{2\eps}g_s^2)^2\,
    \frac{16z_3^2}{z_1z_2s_{13}s_{23}}\,  
    C_3^2\, |\cM_{f_3\ldots f_n}|^2\,.
    \label{eq:2soft3collab}
\end{equation}
For the non-abelian part, we obtain
\begin{equation}
    \mathscr{C}_{123} \mathscr{S}_{12}^{gg, (nab)} |\cM_{g_1g_2f_3\ldots f_n}|^2 =
    - (\mu^{2\eps}g_s^2)^2\, C_A\,
    U_{33;12}^{gg (nab)}\,
    C_3\, |\cM_{f_3\ldots f_n}|^2\,,
    \label{eq:2soft3collnab}
\end{equation}
where
\begin{equation}
    U_{33;12}^{gg (nab)} = 
    8\, \left( \frac{s_{13}s_{23}}{s_{3(12)}^2}
    + \frac{z_1z_2}{(z_1+z_2)^2} \right) \frac{1-\epsilon}{s_{12}^2}
    - 2\cS_{3}^{gg}(p_1,p_2) \,,
    \label{eq:U_ggnab2soft3coll}
\end{equation}
with $\cS_{3}^{gg}(p_1,p_2)$ given by eq.~(\ref{eq:csieq}) with $i=3$.
Equation~(\ref{eq:2soft3collab}) and (\ref{eq:2soft3collnab}) agree with the corresponding overlap of the double soft gluon with the triple collinear limit of ref.~\cite{Somogyi:2005xz}. Equivalently, eqs.~\eqref{eq:2soft3collab} and~\eqref{eq:2soft3collnab} can be cast in the form of a double soft limit of a triple collinear splitting amplitude,
\beq\bsp
\mathscr{S}_{12}^{gg}&
\left[\left(\frac{2\mu^{2\eps}g_s^2}{s_{1\ldots m}}\right)^{m-1}\hat{P}_{g_1g_2f_3}^{ss'}\right] =  (\mu^{2\eps}g_s^2)^2\,C_3\,\left[
    \frac{16z_3^2}{z_1z_2s_{13}s_{23}}\,  
    C_3- \, C_A\,
    U_{33;12}^{gg (nab)}\right]\, \delta^{ss'}\,.
    \esp\eeq

\subsection{Double soft gluon limit of a quadruple collinear limit}
\label{sec:softggtoggff}

In the case of the double soft gluon emission out of a quadruple collinear limit, the non-abelian part follows,
similar to Sec.~\ref{sec:softqtoggq}, the guidelines of
single soft gluon emission of a triple collinear limit,
as well as of soft $q\bar{q}$ pair emission of a quadruple collinear limit of Sec.~\ref{sec:softqqtoqqff}.
We denote the two soft gluons as partons 1 and 2 within the collinear limit of the process $f\to g_1 g_2 f_3 f_4$.
The functional form of eq.~(\ref{eq:P_s1_c123})
is characteristic of soft emissions which give rise to two-parton
colour correlations within a collinear structure made of two hard partons. 
Thus, it applies to the non-abelian part
of the double soft gluon emission
within a quadruple collinear amplitude, and we can recycle
eq.~(\ref{eq:qqtoffqq}), up to
replacing the factors 
$U_{jk;12}^{q\bar{q}}$ in eq.~(\ref{eq:qqdoublesoft_coeffu}) with 
$U_{jk;12}^{gg (nab)}$ in eq.~(\ref{eq:U_ggnabdoublesoft}),
\begin{align}\label{eq:ggtoffggnab}
&\mathscr{S}_{12}^{gg (nab)}\left[\left(\frac{2\mu^{2\eps}g_s^2}{s_{1234}}\right)^{3}\hat{P}_{g_1g_2f_3f_4}^{ss'}\right] \\
\nonumber& =- (\mu^{2\eps}g_s^2)^2\, C_A\, \left[ U_{33;12}^{gg (nab)} {\bf T}_3^2
+ U_{34;12}^{gg (nab)} ({\bf T}_P^2-{\bf T}_3^2-{\bf T}_4^2) + U_{44;12}^{gg (nab)} {\bf T}_4^2 \right]
\left(\frac{2\mu^{2\eps}g_s^2}{s_{34}}\right)\hat{P}_{f_3f_4}^{ss'}\,,
\end{align}
where ${\bf T}_P^2=C_P$ denotes the colour coefficient of the parent $P$ of the collinear set.

For the abelian part, we spell out the quadruple sum in
eq.~(\ref{eq:P_ggdoublesoft_ab}) with $3\le i,j,k,\ell\le 4$,
and after using colour conservation,
we obtain
\beq\bsp
&\mathscr{S}_{12}^{gg (ab)}\left[\left(\frac{2\mu^{2\eps}g_s^2}{s_{1234}}\right)^{3}\hat{P}_{g_1g_2f_3f_4}^{ss'}\right] \\
&\quad = (\mu^{2\eps}g_s^2)^2\,
\left[ U_{33;1} {\bf T}_3^2
+ U_{34;1} ({\bf T}_P^2-{\bf T}_3^2-{\bf T}_4^2) + U_{44;1} {\bf T}_4^2 \right] \\
&\qquad\qquad\times \left[ U_{33;2} {\bf T}_3^2
+ U_{34;2} ({\bf T}_P^2-{\bf T}_3^2-{\bf T}_4^2) + U_{44;2} {\bf T}_4^2 \right]\,
\left(\frac{2\mu^{2\eps}g_s^2}{s_{34}}\right)\hat{P}_{f_3f_4}^{ss'}\,,
\label{eq:ggtoffggab}
\esp\eeq
with the coefficients $U_{jk;l}$ defined in eq.~(\ref{eq:ucoeff}).

The double soft gluon emission
within a quadruple collinear amplitude is given by the
sum of eqs.~(\ref{eq:ggtoffggnab}) and (\ref{eq:ggtoffggab}),
\begin{equation}
    \mathscr{S}_{12}^{gg}\left[\left(\frac{2\mu^{2\eps}g_s^2}{s_{1234}}\right)^{3}\hat{P}_{g_1g_2f_3f_4}^{ss'}\right] =
    \left( \mathscr{S}_{12}^{gg (ab)} + \mathscr{S}_{12}^{gg (nab)} \right)
    \left[\left(\frac{2\mu^{2\eps}g_s^2}{s_{1234}}\right)^{3}\hat{P}_{g_1g_2f_3f_4}^{ss'}\right]\,.
    \label{eq:ggtoffgg}
\end{equation}

As in Sec.~\ref{sec:softqqtoqqff}, through eq.~(\ref{eq:ggtoffgg}) we may describe
three processes, whose colour algebra is the same as in
eq.~(\ref{eq:twohardalg}),
\beq\bsp
&q\to g_1 g_2 g_3 q_4\,, \qquad
{\bf T}_3^2=C_A\,,\quad {\bf T}_4^2={\bf T}_P^2=C_F\,, \\
&g\to g_1 g_2 q_3 \bar{q}_4\,, \qquad
{\bf T}_3^2={\bf T}_4^2=C_F\,,\quad {\bf T}_P^2=C_A\,, \\
&g\to g_1 g_2 g_3 g_4\,, \qquad
{\bf T}_3^2={\bf T}_4^2={\bf T}_P^2=C_A\,.
\label{eq:twohardblg}
\esp\eeq
We have checked that, using the color factors in the first line of eq.~\eqref{eq:twohardblg}, then our results for the splitting amplitude $\hat{P}_{gggq}^{ss'}$ are consistent with the soft factorisation in eq.~\eqref{eq:ggtoffgg}.

\section{Conclusions}
\label{sec:conclusion}

In this paper, we have computed
the quadruple-collinear splitting amplitudes for a quark parent in CDR. These can be found in computer-readable form in the ancillary files~\cite{QuadColKernelsWebsite}. Further, we have considered
the iterated limit when $m'$ massless partons become collinear to each other within a bigger set of $m$ collinear partons,
specifying it to the cases
when two or three partons become collinear to each other within a set of four collinear partons.
Likewise, we have analysed the iterated limits when one gluon or a $q\bar{q}$ pair or two gluons become soft within a set of $m$ collinear partons, specifying then the cases when $m\le 4$. 

Our results provide another important building block to understand the universal infrared structure of QCD amplitudes at N$^3$LO, which is a cornerstone to construct a substraction method at this order. However, more developments are needed before the complete structure of infrared divergences at N$^3$LO is known. Currently we know the structure of infrared singularities of massless amplitudes with up to three loops~\cite{Catani:1998bh,Sterman:2002qn,Becher:2009cu,Gardi:2009qi,Almelid:2015jia,Almelid:2017qju}. Soft singularities are known for the emission of up to three particles at tree-level~\cite{Bassetto:1984ik,Proceedings:1992fla,Catani:1999ss,Catani:2019nqv} and for the emission of a single soft gluon at one and two loops~\cite{Bern:1998sc,Bern:1999ry,Catani:2000pi,Duhr:2013msa,Li:2013lsa}.\footnote{The two-loop soft-current of Refs.~\cite{Duhr:2013msa,Li:2013lsa} is only valid for amplitudes with two hard partons.} Collinear splitting amplitudes are known at tree-level and one-loop for the emission of up to three particles~\cite{Campbell:1997hg,Catani:1999ss,DelDuca:1999iql,Kosower:2002su,Bern:1994zx,Bern:1998sc,Kosower:1999rx,Bern:1999ry,Catani:2003vu,Badger:2015cxa}\footnote{But for the one-loop collinear splitting amplitude $q\to ggq$, which at present is unknown.} and at two-loops for two collinear partons~\cite{Bern:2004cz,Badger:2004uk,Duhr:2014nda}. In this paper we have added to this list the tree-level quadruple-collinear splitting amplitudes for a quark parent in CDR, which so far had only been known in four dimensions for fixed external helicities~\cite{DelDuca:1999iql,Birthwright:2005ak,Duhr:2006}. What still needs to be examined in order to fully understand the universal infrared structure of massless QCD amplitudes at N$^3$LO are therefore the one-loop soft current for the emission of a pair of soft partons as well as the tree-level quadruple-collinear splitting amplitudes for a gluon parent. The latter will be provided in a forthcoming paper.

    \section*{Acknowledgements}
    The authors are grateful to Fabrizio Caola and Lorenzo Tancredi for discussions. CD and MM acknowledge the hospitality of the ETH Zurich, and MM also acknowledges the hospitality of the TH Department of CERN, at various stages of this work.
    This work was supported in part by the ERC starting grant  637019 ``MathAm'' (CD), the FRIA grant of the Fonds National de la Recherche Scientifique (FNRS), Belgium (MM), the European Research Council (ERC) under grant agreement No 694712 (PertQCD) and the Swiss National Science Foundation (SNF) under contract agreement No 179016 (VD, RH, AL).


	\appendix
	
	\section{Kinematics of the collinear limit}
	\label{app:boost}

We use a generic light-cone decomposition of $m$ massless momenta,
\begin{equation} 
\label{eq:collinear_app1}
p_i^\mu = x_i\pM{\mu} + k_{\perp i}^\mu + a_i n^\mu\,, \qquad i=1,\ldots,m\,,
\end{equation}
where the light-like momentum,
\begin{equation}
\pM{} = ( \sqrt{ ({\bf p}_1+\ldots + {\bf p}_m)^2}, {\bf p}_1+\ldots + {\bf p}_m )\,,
\label{eq:collinear_app2}
\end{equation}
specifies a light-cone direction, $\pM{} \cdot k_{\perp i} = 0$,
$n^\mu$ is an auxiliary light-like vector, which specifies how that light-cone direction is approached, $n\cdot k_{\perp i} = 0$,
and $x_i$ are the longitudinal momentum fractions with respect to the total momentum $P^\mu = \sum_{i=1}^m p_i^\mu$.
Note that
\begin{equation} 
\label{eq:collinear_app3}
x_i = \frac{p_i\cdot n}{\pM{} \cdot n}\,, \qquad  a_i = \frac{p_i\cdot \pM{}}{\pM{}\cdot n }\,,
\qquad i=1,\ldots,m\,,
\end{equation}
and the on-shellness condition, $p_i^2 = 0$, allows us to fix
\begin{equation} 
\label{eq:collinear_app4}
a_i = - \frac{k_{\perp i}^2}{2x_i \pM{}\cdot n}\,,
\end{equation}
so we arrive at the usual expression in eq.~(\ref{eq:collinear_momenta}) for the light-cone decomposition.

Note that eq.~(\ref{eq:collinear_app3}) implies that
\begin{equation} 
\label{eq:collinear_app5}
\sum_{i=1}^m x_i = \frac{P\cdot n}{\pM{} \cdot n} \ne 1\,,
\end{equation}
and
\begin{equation} 
\label{eq:collinear_app6}
\sum_{i=1}^m p_i^\mu = P^\mu = \frac{P\cdot n}{\pM{} \cdot n} \pM{\mu} + 
\sum_{i=1}^m k_{\perp i}^\mu + \frac{P\cdot \pM{}}{\pM{} \cdot n} n^\mu\,,
\end{equation}
which shows that in general $\sum_{i=1}^m k_{\perp i}^\mu \ne 0$.
Only in the strict collinear limit ${f \to f_1\ldots f_m}$, for which $P^\mu\to \pM{\mu}$, are the constraints
$\sum_{i=1}^m x_i =1 $ and $\sum_{i=1}^m \K{i}^\mu = 0$ fulfilled.

However, the longitudinal-boost invariant variables,
\begin{equation}
z_i = \frac{x_i}{\sum_{j=1}^m x_j} = \frac{p_i\cdot n}{P\cdot n}\,, 
\qquad 
\KT{i}^\mu = \K{i}^{\mu}- z_i\sum_{j=1}^m \K{j}^{\mu}\,,
\label{eq:boostapp}
\end{equation}
satisfy the constraints, $\sum_{i=1}^m z_i =1$ and $\sum_{i=1}^m\KT{i}^\mu= 0$ also away from the strict collinear limit.

\section{Kinematics of the strongly-ordered collinear limit}
\label{app:nested}
The light-cone decomposition in eq.~(\ref{eq:collinear_momenta}) may be performed with respect to a direction specified by a light-like momentum based
on any $m'$-parton subset of the $m$ collinear partons,\footnote{Without loss of generality, we assume the subset to be made of the first $m'$ partons with momenta $\{p_1,\ldots, p_{m'}\}$.}
\begin{align}
    p_i^\mu = y_i\pN{\mu} + \Kb{i}^\mu - \frac{\Kb{i}^2}{2y_i}\frac{n^\mu}{\pN{} \cdot n},
\end{align}
where $y_i$ are the longitudinal momentum fractions with respect to $P'^{\mu} = \sum_{i=1}^{m'} p_i^\mu$, and 
\begin{equation} 
\label{eq:collinear_app7}
\pN{} = ( \sqrt{ ({\bf p}_1+\ldots + {\bf p}_{m'})^2}, {\bf p}_1+\ldots + {\bf p}_{m'} )\,, \qquad m'\le m\,.
\end{equation}
Similarly, one may choose the longitudinal-boost invariant quantities
\begin{align}
\zeta_i = \frac{y_i}{\sum_{j=1}^{m'} y_j} = \frac{p_i\cdot n}{P'\cdot n}\,, 
\qquad 
\KTb{i}^\mu = \Kb{i}^{\mu}- \zeta_i\sum_{j=1}^{m'} \Kb{j}^{\mu}\,,
\label{eq:boost_subsetapp}
\end{align}
which satisfy the constraints $\sum_{i=1}^{m'} \zeta_i =1$ and $\sum_{i=1}^{m'}\KTb{i}^\mu= 0$ away from the strict collinear limit 
$P' + p_{m'+1}+\ldots +p_m \to \pM{\mu}$.
By decomposing the momenta $p_i$ with respect to the light-cone directions $\pM{\mu}$ and $\pN{\mu}$, it is possible to connect longitudinal-boost invariant quantities in those directions,
\begin{equation}
\zeta_i = \frac{p_i \cdot n}{P' \cdot n} =
\frac{p_i \cdot n / P \cdot n}{\sum_{j=1}^{m'} p_j \cdot n / P \cdot n} =
\frac{z_i}{\sum_{j=1}^{m'} z_j},
\quad i=1,\ldots,m'\,.
\label{eq:connect_pfrac}
\end{equation}
For instance, the triple collinear limit involves a $(1\to 2)$ subcurrent and one may define longitudinal-boost invariant quantities of the $(12)$ direction,
\begin{align}
    \pN{} = ( \sqrt{ ({\bf p}_1+ {\bf p}_{2})^2}, {\bf p}_1+ {\bf p}_{2} )\,,
\end{align}
by
\begin{equation}
\zeta_i = \frac{p_i\cdot n}{p_{12} \cdot  n} \equiv \frac{z_i}{z_1+z_2},
\qquad 
\KTb{i}^\mu = \K{i}^{\mu} - \zeta_i (\K{1}+\K{2})^{\mu}, \quad i=1,2 \,.
\end{equation}
%

Next, we consider the case where the light-like vector $\pN{}$ describes the light-cone direction of an $m'$-parton collinear subset.
We may perform a generic Sudakov decomposition of $\pN{}$ with respect to the light-like direction $\pM{}$ as follows:
%
%
\begin{equation} \label{eq:sub_collinear_lightcone}
    \pN{\mu} = \alpha \pM{\mu} + K^\mu - \frac{K^2}{2\alpha}\frac{n^\mu}{n\cdot \pM{}},
\end{equation}
where $K = \sum_{l=1}^{m'} \K{l}$. The coefficient $\alpha$ is fixed by the transversality condition $K \cdot n = 0$,
\begin{align}
\label{eq:alpha}
\alpha &= \frac{\pN{}\cdot n}{\pM{}\cdot n} 
=\frac{x_i}{y_i}, \quad \forall 1 \leq i \leq m'\,.
\end{align}
We note that {\it a priori} both $K$ and $\alpha$ depend on variables that do not satisfy the constraints, $\sum_{i=1}^m x_i \neq 1$, $\sum_{i=1}^{m'} y_i \neq 1$ and $\sum_{i=1}^m \K{i} \neq 0$.
However, following the discussion in Sec.~\ref{sec:setup}, we can define
longitudinal-boost invariant quantities by (cf.~eqs.~(\ref{eq:boost}) and~(\ref{eq:boost_subset})),
\begin{align}
    \tilde{\alpha} 
    &= \frac{z_i}{\zeta_i} , \quad \forall 1 \leq i \leq m', \\
    \quad
    \widetilde{K}^\mu &= \sum_{l=1}^{m'} \KT{l}^\mu = \sum_{l=1}^{m'}\left( \K{l}^\mu -  z_l \sum_{j=1}^{m} \K{j}^\mu \right )\,,
\end{align}
such that $\widetilde{K}^\mu= - \sum_{i=m'+1}^m\K{i}^\mu$ and $\tilde{\alpha} = \sum_{i=1}^{m'} z_i$. In particular, the longitudinal-boost invariant momentum fractions $\zeta_i$ and $z_i$ satisfy eq.~\eqref{eq:connect_pfrac} also in the strongly-ordered limit.
To avoid cluttering notation, we implicitly assume longitudinal-boost invariant quantities and drop the tilde.

Substituting eq. \eqref{eq:sub_collinear_lightcone} into eq. \eqref{eq:Sudakov1.2} with $n' = n$, we cast the light-cone parametrisation of the $m'$-parton subset into an auspicious form,
\begin{equation}
p^\mu_i = \zeta_i \alpha \pM{\mu} + \left(\zeta_i K^\mu + \Kb{i}^\mu \right) 
- \left( 
\zeta_i \frac{K^2}{2\alpha}\frac{n^\mu}{ n\cdot \pM{}} + 
\frac{ \Kb{i}^2}{2 \zeta_i}\frac{n^\mu}{n \cdot \pN{}} 
\right), 
\quad
i =1,\ldots,m'\,.
\label{eq:Sudakov2}
\end{equation}
Contracting both sides of eqs. \eqref{eq:collinear_momenta} and \eqref{eq:Sudakov2} with $\K{i}$, one obtains a relation between the transverse components, modulo gauge terms $\pM{}$ and $n$ that are perpendicular to $\K{i}$,
\begin{align}
\K{i}^2 &= \zeta_i K \cdot \K{i} + \Kb{i} \cdot \K{i} \\
\Rightarrow \K{i}^\mu &= \zeta_i K^\mu + \Kb{i}^\mu + a \pM{\mu} + b n^\mu\,.
\label{perp_relation}
\end{align}
The longitudinal components are negligible in the strict collinear limit $\pN{} || \pM{}$, since the transverse momenta trivially line up. Therefore, eq. \eqref{perp_relation} is a statement about how the transverse components of each collinear set approach the strongly-ordered limit relative to each other.
The coefficients $a$ and $b$ are fixed by on-shellness, $\pM{2}=0$ and $n^2 =0$, and the transversality conditions, $\K{i} \cdot n = \Kb{i} \cdot n = K \cdot n = 0$ and $K \cdot \pM{} = 0$,
\begin{align}
a= 0, \quad b = - \frac{\Kb{i} \cdot \pM{}}{n \cdot \pM{}}\,.
\end{align} 
We can eliminate $\Kb{i} \cdot \pM{}$ by contracting eq. \eqref{eq:sub_collinear_lightcone} with $\Kb{i}$ and using $\pN{}\cdot \Kb{i} =0$. Then we obtain
\begin{equation}
\Kb{i} \cdot \pM{} = -\frac{1}{\alpha}K\cdot \Kb{i} \neq 0\,,
\label{eq:Kb_ptildeM}
\end{equation}
and the relation between the transverse momenta reads
\begin{equation}
\K{i}^\mu = \zeta_i K^\mu + \Kb{i}^\mu +  \frac{K\cdot \Kb{i}}{\alpha}\frac{ n^\mu}{ n \cdot \pM{}}\,.
\end{equation}
Therefore, the transverse momenta belonging to the $m'$-parton collinear subset are not orthogonal to the $m$-parton collinear direction, except in the strict collinear limit. Equation~\eqref{eq:Kb_ptildeM} is required when computing collinear limits of the splitting amplitudes themselves. This procedure is outlined in App.~\ref{app:iterated_limit_of_split}.

\section{The iterated collinear limit}
\label{app:iterated_limit_of_split}
From eq.~\eqref{eq:P_coll} we construct a quantity called the strongly-ordered amplitude,
\begin{equation}
\hat{P}_{f_1\ldots f_m}^{\text{s.o.}\,;ss'} = 
\hat{P}^{hh'}_{f_1\ldots f_{m'}}\,
\hat{H}^{hh';ss'}_{f_{(1\ldots m')}f_{m'+1}\ldots f_m}\,,
\end{equation}
which depends on the quantum numbers and light-cone kinematics of both the $m$-parton collinear set and it's $m'$-parton subset. It is obtained by summing over the helicities $(h,h')$ of the parent parton of the collinear subset. In case the parent with helicities $(s,s')$ is a quark, the strongly-ordered amplitude reduces to
\begin{align}
    \hat{P}_{f_1\ldots f_m}^{\text{s.o.}\,;ss'} = \delta^{ss'} \hat{P}_{f_1\ldots f_m}^{\text{s.o.}}\,,
\end{align}
due to eqs.~\eqref{eq:P_coll_unpol} and \eqref{eq:H_parentgluon}. The strongly-ordered splitting amplitude can be obtained by performing the $m'$-parton iterated limit on the $m$-parton splitting amplitude,
\begin{equation}
 \hat{P}_{f_1\ldots f_m}^{\text{s.o.}\,;ss'}
 =
 \left(\frac{s_{1\ldots m'}}{s_{[1\ldots m']\ldots m}}\right)^{m'-1}
\mathscr{C}_{1\ldots m'} \hat{P}^{ss'}_{f_1\ldots f_m}
\,.
\label{eq:so_from_split}
\end{equation} 
As a first step, performing the limit on the right-hand side requires the change of variables,
\begin{align}
    z_i = \zeta_i\left(1-\sum_{j=m'+1}^m z_j\right), \quad  1\leq i\leq m'\,,
\end{align}
which separates the kinematics of the lower-order $m'$-parton splitting process from that of the remaining $(m-m')$ partons with momenta $\{p_{m'+1},\ldots,p_m\}$. Next, we perform a light-cone decomposition of the sub-energies,
\begin{align}
\label{eq:so_sij_param2_1}
s_{ij} &\to \lambda'^2 s_{ij},\quad 1\leq i,j \leq m', \\
s_{ir} &\to 2p_r \cdot \left( \zeta_i \pN{} + \lambda'\Kb{i} - \lambda'^2\frac{ \Kb{i}^2}{2\zeta_i}\frac{n}{n\cdot\pN{}} \right),
\quad 1\leq i\leq m', \quad m'+1 \leq r \leq m,
\label{eq:so_sij_param2_2}
\end{align}
applying a uniform rescaling $\Kb{i} \to \lambda' \Kb{i}$.
To complete the Sudakov expansion of \eqref{eq:so_sij_param2_2}, we use eq. \eqref{eq:Kb_ptildeM} to obtain the non-trivial relation,
\begin{equation}
p_r \cdot \Kb{i} = \K{r} \cdot \Kb{i} - \frac{z_r}{\alpha} K \cdot \Kb{i},\quad 1\leq i\leq m', \quad m'+1 \leq r \leq m,
\end{equation}
where $K = \sum_{l=1}^{m'} \K{l}$. Therefore, the full substitution reads
\beq\bsp
s_{ir} \to& \quad
\zeta_i s_{[1\ldots m']r} +
\lambda'  \left( 2\K{r} \cdot \Kb{i} - \frac{2z_r}{\alpha}K \cdot \Kb{i}\right)
- \lambda'^2 \frac{ z_r}{\alpha \zeta_i}\Kb{i}^2, \\
&\quad 1\leq i\leq m', \quad m'+1 \leq r \leq m, 
\esp\eeq
where $s_{[1\ldots m']r} = 2p_r \cdot \pN{}$. Finally, the strongly-ordered splitting amplitude is obtained by series expanding in $\lambda'$ and keeping the leading divergent terms only. This way, we were able to verify all the strongly-ordered limits of the quadruple-collinear splitting amplitudes, by exploiting the equivalence in eq.~\eqref{eq:so_equivalence}.

\section{The three-parton splitting tensor $H_{g_{(12)}g_3q_4}$}
\label{app:3partonhelicity_tensor}
In this section we provide the results for the three-parton splitting tensor $H_{g_{(12)}g_3q_4}$ defined in eq.~(\ref{eq:Hggq}), in terms of the tensor structure of eq.~\eqref{eq:helicity_tensor}. We added a subscript $(12)$ to denote the on-shell momentum of the gluon sub-parent. Furthermore, we define the shorthand
\begin{equation}
    z_{1\ldots j} = z_1 + \ldots + z_j\,, \qquad
    \bar{z}_i = 1- z_i\,, \qquad
    \K{1\ldots j} = \K{1} + \ldots + \K{j}\,.
\end{equation}
In what follows, we have eliminated $z_{12}$ and $\K{12}$ using the constraints. The sub-energies $s_{[ij]k}$ are defined in eq.~\eqref{eq:s_square}. The coefficients in eq.~\eqref{eq:helicity_tensor} belonging to the `abelian' pieces of $H_{g_{(12)}g_3q_4}$ are given by
\begingroup
\allowdisplaybreaks
\begin{align}
A^{\text{(ab)}}_{g_{(12)} g_3 q_4} &= \frac{1}{2 z_3} 
\Bigg\{
\frac{s_{[12]3}^2 (2(\bar{z}_4+z_3)-(D-2)z_3)z_4}{2s_{34} s_{[12]4}} 
- \frac{s_{[12]4}(z_3(1+z_{34}) -2z_4\bar{z}_4)}{s_{34}}\nonumber\\
&+ \frac{s_{34}(\bar{z}_3^2 - z_4^2)}{s_{[12]4}}
+ s_{[12]3}\left(\frac{3z_4\bar{z}_4 -(1+z_3)z_3 }{s_{34} } 
+ \frac{ \bar{z}_3^2 + (1+z_3-2z_4)z_4 }{s_{[12]4}} \right) \nonumber \\
&+(2-z_3)z_4 - 3z_4^2 +\bar{z}_3 
+s_{[12]3}z_3 \bar{z}_4(D-2) \frac{1}{2}\left (\frac{1}{s_{34}}+\frac{1}{s_{[12]4}} \right )
\Bigg\}\,, \\
B_{33,g_{(12)} g_3 q_4}^{\text{(ab)}} &=
\frac{s_{[12]34}^2z_4}{s_{34} z_3 (1-z_{34})^2} 
\Bigg\{
\frac{(D-2)(z_3z_4 + \bar{z}_4^2)z_3 + 4z_4}{s_{[12]4}} 
+\frac{z_4s_{34}(4\bar{z}_3 +(D-2)z_3^2)}{s_{[12]4}^2} 
\Bigg\}, \\
B_{44,g_{(12)} g_3 q_4}^{\text{(ab)}} &=
\frac{s_{[12]34}^2\bar{z}_3}{s_{34} z_3 (1-z_{34})^2} 
\left\{
\frac{s_{34}(4\bar{z}_3 + (D-2)z_3^2)\bar{z}_3}{s_{[12]4}^2} 
-\frac{z_3^2(D-2)(z_{34}-2) -4z_4)}{s_{[12]4}}
\right\}, \\
B_{34,g_{(12)} g_3 q_4}^{\text{(ab)}} &\equiv B_{43,g_{(12)} g_3 q_4}^{\text{(ab)}} =
\frac{s_{[12]34}^2}{2 s_{34} z_3 (1-z_{34})^2} 
\Bigg\{
\frac{2s_{34}\bar{z}_3z_4(4\bar{z}_3 + (D-2)z_3^2)}{s_{[12]4}^2} \nonumber\\
&+ \frac{(D-2)\left(\bar{z}_4^2-z_3z_4 \left(2 z_3+2 z_4-5\right)-z_3\right)z_3
+4z_4 \left(\bar{z}_3+z_4\right)}{s_{[12]4}}
\Bigg\},
\end{align}
while the `non-abelian' coefficients read
\begin{align}
A^{\text{(nab)}}_{g_{(12)} g_3 q_4} &=
\frac{s_{[12]4}^2}{\bar{z}_4s_{[12]3}} 
\left(\frac{z_3^2}{\bar{z}_4 s_{[12]3}}-\frac{z_3\bar{z}_4-2 \bar{z}_4^2-z_3^2}{4 s_{3,4}}\right)
-\frac{s_{3,4}s_{[12]4}}{ s_{[12]3}^2}\frac{2 z_3 \left(1-z_{34}\right) }{\bar{z}_4^2}
\nonumber \\
&+\frac{s_{3,4}^2\left(1-z_{34}\right)^2}{\bar{z}_4s_{[12]3}} 
\bigg(\frac{\left(z_3+\bar{z}_4\right)}{4 z_3  s_{[12]4}}
+\frac{1}{\bar{z}_4 s_{[12]3}}\bigg)
-\frac{ s_{3,4} \left(1-z_{34}\right)}{4 s_{[12]3}}
\bigg( \frac{z_3 \left(z_4+7\right)}{\bar{z}_4^2} \nonumber \\
&-\frac{3\bar{z}_4}{z_3}- \frac{4}{\bar{z}_4}\bigg)
-\frac{s_{3,4}\left(1-z_{34}\right)}{8 s_{[12]4}} 
\bigg((D-2)
+\frac{2(z_3+\bar{z}_4)(z_3+3z_4-2)}{ z_3\bar{z}_4}\bigg)
\nonumber \\
&+\frac{s_{[12]3}^2}{8s_{34}s_{[12]4}}
\left(  z_4(D-2)-  \frac{2z_4(z_3 +\bar{z}_4)}{z_3}\right)
-\frac{s_{[12]3}}{8s_{[12]4}}
\bigg((D-2) \left(\bar{z}_3-2 z_4\right) \nonumber \\
&+2z_4 + \frac{2z_3(1-2z_4)}{\bar{z}_4}+2\frac{ z_4\left(4-3 z_4\right)-1}{z_3}\bigg) 
-\frac{s_{[12]3}}{8s_{3,4}}
\bigg((D-2) \left(z_3-z_4\right) \nonumber \\
&-\frac{2z_3(1-2z_4)}{\bar{z}_4} + \frac{2\bar{z}_4(3z_4-z_3)}{z_3}\bigg)
+\frac{s_{[12]4}}{8 s_{3,4}}
\bigg(2-2(z_4-z_3)\frac{2\bar{z}_4^2 + z_3^2}{z_3 \bar{z}_4}
\nonumber \\
&-(D-2)z_3\bigg)
+\frac{s_{[12]4}}{4s_{[12]3}}
\left( \frac{z_3(4z_3+(z_3 -\bar{z}_4)(z_4+3))}{\bar{z}_4^2} 
+ \frac{2\bar{z}_4^2}{z_3}\right)
   \nonumber \\
&+\frac{1}{8}(D-2)\left(1+z_4\right)
+\frac{1}{4} \left(2 z_4-1\right)\left(1-\frac{3\bar{z}_4}{z_3} \right)
-\frac{ z_3 \left(1+z_4\right) \left(1 - z_{34}\right)}{2\bar{z}_4^2}\\
B_{33,g_{(12)} g_3 q_4}^{\text{(nab)}} &= 
\frac{s_{[12]34}^2}{2(1-z_{34})^2}
\Bigg\{
\frac{2\bar{z}_4(4z_4 + (D-2)\bar{z}_4^2)}{s_{[12]3}^2} 
- \frac{4z_4(z_3-2\bar{z}_4)}{s_{34}s_{[12]3}\bar{z}_4} 
\nonumber\\
&- \frac{z_4(4\bar{z}_3 + \bar{z}_4^2(D-2)(\bar{z}_4 + z_3) + 4(1-2z_4)z_4)}{s_{[12]3}s_{[12]4}\bar{z}_4}
- \frac{4 z_4^2}{s_{34}s_{[12]4}z_3}
\nonumber\\
&+\left(\frac{\bar{z}_4}{s_{[12]3}} - \frac{z_4}{s_{[12]4}} \right)
\frac{(D-2)(z_4(z_{34}-2)+1)}{s_{34}}
\Bigg\}, \\
B_{44,g_{(12)} g_3 q_4}^{\text{(nab)}} &= 
\frac{s_{[12]34}^2}{2(1-z_{34})^2}
\Bigg\{
 \frac{2z_3^2 ( 4z_4 + (D-2)\bar{z}_4^2 )}{s_{[12]3}^2 \bar{z}_4}
-\frac{\bar{z}_3 ( 4z_4 + (D-2)(2-z_{34})z_3^2 )}{s_{34}s_{[12]4}z_3} 
 \nonumber\\
&+ \frac{\bar{z}_3}{s_{[12]3}s_{[12]4}} \left(
\frac{4  \left(z_3 \left(2 z_3-3\right)+\bar{z}_4\right)}{z_3}
-(D-2)(\bar{z}_4 + z_3) z_3 
\right)
\nonumber\\
&+\frac{1}{s_{34}s_{[12]3}} \left(
4 + 4z_3 \left( \frac{z_3}{\bar{z}_4}-2 \right)
+ (D-2)(2-z_{34})z_3^2
\right)
\Bigg\}, \\
B_{34,g_{(12)} g_3 q_4}^{\text{(nab)}} &\equiv B_{43,g_{(12)} g_3 q_4}^{\text{(nab)}} =
\frac{s_{[12]34}^2}{4(1-z_{34})^2}
\Bigg\{
\frac{4z_3 ( \bar{z}_4^2(D-2) + 4z_4 )}{s_{[12]3}^2} \nonumber\\
&- \frac{
4z_4 (z_4 + \bar{z}_3) + z_3(D-2)(z_4(2z_3-1)(2-z_{34}) + \bar{z}_3)}{z_3 s_{34}s_{[12]4}} 
\nonumber\\
&+ \frac{4\bar{z}_4^2 z_4 - \bar{z}_4 z_3 (4 + (D-2)\bar{z}_4^2 + 20z_4 )
+ 2z_3^2( z_4(6 - 8z_4 - (D-2)\bar{z}_4^2) +4 )}{z_3 \bar{z}_4 s_{[12]3}s_{[12]4}} \nonumber\\ 
&- \frac{ 
z_3(8+ \bar{z}_4^2(D-2)(2z_4-3) -4z_4) -4\bar{z}_4(1+2z_4)}{\bar{z}_4s_{34}s_{[12]3}} \nonumber\\
&- \frac{z_3^2(s_{34}-s_{[12]4})( \bar{z}_4(D-2)(2z_4-1) + 4  )}{\bar{z}_4s_{34}s_{[12]3}s_{[12]4}}
\Bigg\}.
\end{align}
\endgroup

\newpage

\bibliographystyle{JHEP}
\bibliography{quadri}
\end{document}